\documentclass[preprint,showpacs,preprintnumbers,amsmath,amssymb, superscriptaddress]{revtex4}

\usepackage{textcomp}
\usepackage{makeidx}
\usepackage{amsmath}
\usepackage{subfigure}
\usepackage{amssymb}
\usepackage{hyperref}
\usepackage{cleveref}
\usepackage{graphicx}
\usepackage[utf8]{inputenc}
\usepackage{float}
\usepackage{color}
\usepackage{epsfig}
\usepackage{eucal}
\usepackage{amsmath}
\usepackage{tabularx}
\usepackage{tabulary}

\begin{document}

\title{Charged anisotropic compact objects by gravitational decoupling}

\author{E. Morales}
\email{emc032@alumnos.ucn.cl}
\affiliation{Departamento de F\'isica, Universidad Cat\'olica del Norte, Av. Angamos $0610$, Antofagasta, Chile.}
\author{Francisco Tello-Ortiz}
\email{francisco.tello@ua.cl}
\affiliation{Departamento de F\'isica, Facultad de ciencias básicas, Universidad de Antofagasta, Casilla 170, Antofagasta, Chile.}


\begin{abstract}
In the present article, we have constructed a static charged anisotropic compact
star model of Einstein field equations for a spherically symmetric space-time geometry. Specifically, we have extended the charged isotropic Heintzmann solution to an anisotropic domain. To address this work, we have employed the gravitational decoupling through the so called minimal geometric deformation approach. The charged anisotropic model is representing the realistic compact objects such as $RXJ1856-37$ and $SAX J1808.4-3658(SS2)$. We
have reported our results in details for the compact star $RXJ1856-37$ on the ground of
physical properties such as pressure, density, velocity of sound, energy conditions, stability conditions, Tolman-Oppenheimer-Volkoff equation 
and redshift etc. 
\end{abstract}


\maketitle

\section{Introduction}

Since the birth of the Einstein gravity theory, \emph{general relativity} (GR). It has been a great challenge to find solutions that describe a well behaved structures from the physical point of view in the Universe. The first who gives an exact solution to Einstein field equations describing the exterior of a spherically symmetric and static fluid sphere was K. Schwarzschild \citep{schwar}. Then R. Tolman found several solutions corresponding to a perfect fluid matter distributions \cite{tolman}, but was G. Lamaitre who pointed out that all the structures inside the Universe may contain anisotropic matter distributions, explaining that the spherically symmetry do not require the isotropic condition $p_{r}=p_{t}$ at all \cite{lamei}. On the other hand, the work of Bowers and Liang, about local anisotropic equation of state for relativistic spheres \cite{bowers}, allowed a better understanding respect to this type of matter distributions. Also the studies of Ruderman about more realistic stellar models show that the nuclear matter may be anisotropic at least in certain very high density ranges $(\rho>10^{15} g/cm^{3})$, where the nuclear interactions must be treated relativistically \cite{ruderman}. In the recent years several works available in the literature \cite{kileba,maurya2,hassan,bhar,bhar1,maurya3} (and reference contained therein) address this issue in order to examined how anisotropic matter distribution affects on the effective mass, radius of the stars, central energy density, critical surface redshift and stability of highly compact bodies, since in some cases the presence of anisotropy rises in a repulsive force ($p_{t}>p_{r}$) which counteracts the gravitational gradient \cite{harko1}. Moreover, models with a matter tensor containing anisotropy, must be consistent with physical requirements for astrophysical applications. This is so because the presence of anisotropic pressures leads to values of observed compactness
parameters for several astrophysical bodies \cite{maurya2}.\\
All the mentioned works above, concerned only a neutral spherically symmetric and static configurations. However, it is also interesting study these fluid spheres in presence of a static electric field. As a extension of the exterior Schwarzschild's solution to this context, we have the well known Reissner-Nordstrom solution \cite{reissner,nordstrom}. As was pointed out by Thirukkanesh et al. \cite{Thirukkanesh} it is interesting to note that, in presence of the electromagnetic fields, the collapse of a spherically symmetric matter distribution to a point singularity may be avoided during the gravitational collapse or during an accretion process onto  compact object \cite{krasinski}. In this scenario, the gravitational attraction is counterbalanced by the repulsive Coulomb force in addition to the pressure gradient \cite{bekenstein}. Another important feature is related with the energy density associated to the  electric field, which has a significant role in producing the gravitational mass of the object \cite{deb}. In fact, several literature \cite{monadi,rahaman,varela,negreiros,mello,ray} can be referred to understand the effects of the electric charge on the relativistic compact stellar system. In a more widely context, charged self-gravitating anisotropic fluid spheres have been extensively investigated in general relativity since the pioneering work of Bonnor \cite{bonnor}. In fact, the presence of anisotropy + static electric field enhances the stability and equilibrium conditions of compact objects \cite{maurya3,deb,kileba1,hassan1,maurya4}. Of course, as mentioned earlier each of these ingredients counteracts the gravitational force.\\
So, in the present work we obtain from the charged isotropic Heinzmann's interior solution describing compact star \cite{Thirukkanesh}, an anisotropic extension. It's achieved employing the so ca-
lled \emph{minimal geometric deformation} approach $(MGD)$ \cite{Ovalle,Ovalle9}. This method was originally proposed in the context of
the Randall-Sundrum braneworld \cite{randall,randall1} and was designed to deform the standard Schwarzschild solution \cite{Ovalle4,Ovalle5}. The main point of this scheme is that the isotropic and anisotropic sectors can be split. Therefore, the decoupling of both gravitational sources can be done in a simple  form establishing  a novel way to search new families of anisotropic solutions of Einstein field equations. \\
The paper is organized as follows: Section II presents the Einstein field equations
for an anisotropic matter distributions. In Section III the MGD approach is presented in brief, in order to explain how to generate
arbitrary anisotropic solutions. Section IV is devoted to apply this method to a particular seed solution, the
charged isotropic Heinzmann mo-
del for compact objects. In Section V we analyzed all the requirements for a well behaved solution from the physical point of view. Finally, in section VI we give some conclusions for the reported study.


\section{Main field equations for anisotropic distributions}
The starting point is the static, spherically symmetric
line element represented in Schwarzschild-like coordinates. It reads

\begin{equation}\label{schwarzschild}
ds^{2}=e^{\nu}dt^{2}-e^{\lambda}dr^{2}-r^{2}\left(d\theta^{2}+\sin^{2}\theta d\phi^{2}\right),    
\end{equation}
where $\nu=\nu(r)$ and $\lambda=\lambda(r)$. The metric (\ref{schwarzschild}) is a generic solution of the Einstein field equations
\begin{equation}\label{Einstein}
R_{\mu\nu}-\frac{1}{2}Rg_{\mu\nu}=-\kappa{T}_{\mu\nu}, 
\end{equation}
describing an anisotropic fluid sphere. The coupling constant is given by $\kappa=\frac{8\pi G}{c^{4}}$, from now on we will employ relativistic geometrized units, that is $c=G=1$.\\ 
The stress-energy tensor ${T}_{\mu\nu}$ corresponding to an anisotropic matter distribution, in an orthonormal basis is characterized by ${\rho}$, ${p}_{r}$ and ${p}_{t}$ \cite{Visser}, which are related to the metric functions $\nu$ and $\lambda$ through (\ref{Einstein}). Then the field equations explicitly reads 
\begin{eqnarray}\label{effectivedensity}
8\pi {\rho}&=&\frac{1}{r^2}-e^{-\lambda}\left(\frac{1}{r^2}-\frac{\lambda^{\prime}}{r}\right)\\\label{effectiveradialpressure}
8\pi {p}_{r}&=&-\frac{1}{r^2}+e^{-\lambda}\left(\frac{1}{r^2}-\frac{\nu^{\prime}}{r}\right)\\\label{effectivetangentialpressure}
8\pi {p}_{t}&=&\frac{1}{4}e^{-\lambda}\left(2\nu^{\prime\prime}+\nu^{\prime2}-\lambda^{\prime}\nu^{\prime}+2\frac{\nu^{\prime}-\lambda^{\prime}}{r}\right).
\end{eqnarray}
The primes denote differentiation with respect to $r$. Bianchi identity  invokes the following conservation equation for the stress-energy tensor
\begin{equation}\label{bianchi}
\nabla^{\nu}{T}_{\mu\nu}=0.
\end{equation}
On the other hand we will make use the following representation for the energy-momentum tensor
\begin{equation}\label{effectivestresstensor}
{T}_{\mu\nu}=\tilde{T}_{\mu\nu} + \alpha \theta_{\mu\nu},   
\end{equation}
where the first term in the right hand side represents an isotropic perfect fluid, 
\begin{equation}\label{perfectfluid}
\tilde{T}_{\mu\nu}=\left(\tilde{\rho}+\tilde{p}\right)u_{\mu}u_{\nu}-\tilde{p}g_{\mu\nu}, 
\end{equation}
representing the vector $u^{\mu}=e^{-\nu(r)/2}\delta^{\mu}_{0}$ the unit timelike four-velocity.  Along this work the thermodynamics observable $\tilde{\rho}$ and $\tilde{p}$, correspond to charged isotropic Heintzmann interior solution \cite{Thirukkanesh}. According to this representation, the extra gravitational contribution is given by the $\theta$-term, which causes a deviation from $GR$. In principle this additional gravitational source can be e.g. a scalar field, a vector field or a tensor field. It is coupled to gravity via a dimensionless 
parameter $\alpha$. It noteworthy that in the limit $\alpha\rightarrow0$ $GR$ is recovered, i.e. Einstein equations for isotropic matter distributions are obtained.  \\
In the system of equations (\ref{effectivedensity})-(\ref{effectivetangentialpressure}), 
${\rho}$, ${p}_{r}$ and ${p}_{t}$ represent the effective density, the effective radial pressure and the effective tangential pressure respectively, that are given by
\begin{eqnarray}\label{effecrho}
{\rho}&=&\tilde{\rho}+\alpha \theta^{t}_{t}\\\label{effecpr}
{p}_{r}&=&\tilde{p}-\alpha \theta^{r}_{r}\\ \label{effecpt}
{p}_{t}&=& \tilde{p}-\alpha \theta^{\varphi}_{\varphi}.
\end{eqnarray}
Hence, it is clear that the presence of the $\theta$-term raises an anisotropy if $\theta^{r}_{r}\neq \theta^{\varphi}_{\varphi}$. Thus the effective anisotropy is defined as  
\begin{equation}\label{anisotropy}
\Pi\equiv {p}_{t}-{p}_{r}=\alpha\left(\theta^{r}_{r}- \theta^{\varphi}_{\varphi}\right)  
\end{equation}
Taking into account the expression (\ref{effectivestresstensor}) the corresponding conservation law (\ref{bianchi}) yields to
\begin{equation}
\begin{split}
\tilde{p}^{\prime}+\frac{\nu^{\prime}}{2}\left(\tilde{p}+\tilde{\rho}\right)-\alpha\big[\left(\theta^{r}_{r}\right)^{\prime}  
+\frac{\nu^{\prime}}{2}\left(\theta^{r}_{r}-\theta^{t}_{t}\right)
+\frac{2}{r}\left(\theta^{r}_{r}-\theta^{\varphi}_{\varphi}\right)\big]=0,
\end{split}
\end{equation}
being the above expression a linear combination of the equations (\ref{effectivedensity})-(\ref{effectivetangentialpressure}). To solve the system of equations (\ref{effectivedensity})-(\ref{effectivetangentialpressure}) we will face it applying the $MGD$ scheme \cite{Ovalle}.

\section{Minimal geometric deformation scheme in brief}

Here we present in short the MGD approach, an extensive development of this method is given in references \cite{Ovalle1,Ovalle2,Ovalle3,Ovalle6,Ovalle7,Ovalle8} and recent applications of it can be found in \cite{Gabbanelli,Tello,sharif,Ovalle:2018umz}. So this scheme causes an anisotropic modification to usual solutions of Einstein field equations.  In order to tackle the system of equations (\ref{effectivedensity})-(\ref{effectivetangentialpressure}), we take a spherically symmetric isotropic matter distribution, this is $p_{r}=p_{t}=p$. From this seed solution also are known the metric functions $e^{\lambda}$ and $e^{\nu}$. The output will be a shift in the effective pressures such that $p_{r}\neq p_{t}$. To accomplish it, one makes a most general minimal geometric deformation on the temporal and radial metric functions keeping the spherically symmetry of the original solution   

\begin{eqnarray}\label{deformationnu}
e^{\nu(r)}&\mapsto& e^{\nu(r)}+\alpha h^{*}(r) \\ \label{deformationlambda}
e^{-\lambda(r)}&\mapsto& \mu(r)+\alpha f^{*}(r).
\end{eqnarray}
In the above linear mapping $ h ^ {*} (r) $ and $ f ^ {*} (r) $ are the corresponding deformations. In principle the method allows to us set $ h ^ {*} (r) = 0 $. Therefore all the anisotropic sector $\theta_{\mu\nu}$ relies over the radial deformation (\ref{deformationlambda}). The most remarkable feature of the MGD method  is that it decouple the system (\ref{effectivedensity})-(\ref{effectivetangentialpressure}) resulting in two separated system of equations related only by the metric function $\nu$. One of them corresponds to the standard Einstein equations for the chosen solution (perfect fluid solution), and the second one an effective "quasi-Einstein" system of equations to the anisotropic sector. Then we have

\begin{eqnarray}\label{ro1}
8\pi\tilde{\rho}&=&\frac{1}{r^{2}}-\frac{\mu}{r^{2}}-\frac{\mu^{\prime}}{r}\\\label{p1}
8\pi \tilde{p}&=&-\frac{1}{r^{2}}+\mu\left(\frac{1}{r^{2}}+\frac{\nu^{\prime}}{r}\right)\\\label{p2}
8\pi \tilde{p}&=&\frac{\mu}{4}\left(2\nu^{\prime\prime}+\nu^{\prime2}+2\frac{\nu^{\prime}}{r}\right)+\frac{\mu^{\prime}}{4}\left(\nu^{\prime}+\frac{2}{r}\right),
\end{eqnarray}
along with the conservation equation
\begin{equation}\label{conservde}
\tilde{p}^{\prime}+\frac{\nu^{\prime}}{2}\left(\tilde{\rho} +\tilde{p}\right)=0,    
\end{equation}
this is a linear combination of the equations (\ref{ro1}) -(\ref{p2}). On the other hand we have the following equations to the $\theta$- sector
\begin{eqnarray}\label{cero}
8\pi\theta^{t}_{t}&=&-\frac{f^{*}}{r^{2}}-\frac{f^{*\prime}}{r} 
\\ \label{one}  
8\pi\theta^{r}_{r}&=&-f^{*}\left(\frac{1}{r^{2}}+\frac{\nu^{\prime}}{r}\right)  \\  \label{dos}
8\pi\theta^{\varphi}_{\varphi}&=&-\frac{f^{*}}{4}\left(2\nu^{\prime\prime}+\nu^{\prime2}+2\frac{\nu^{\prime}}{r}\right)-\frac{f^{*\prime}}{4}\left(\nu^{\prime}+\frac{2}{r}\right). \label{tres}
\end{eqnarray}
The corresponding conservation equation $\nabla^{\nu}\theta_{\mu\nu}=0$ then yields to
\begin{equation}\label{conservationtheta}
\left(\theta^{r}_{r}\right)^{\prime}-\frac{\nu^{\prime}}{2}\left(\theta^{t}_{t}-\theta^{r}_{r}\right)-\frac{2}{r}\left(\theta^{\varphi}_{\varphi}-\theta^{r}_{r}\right)=0.  
\end{equation}
Being the equation (\ref{conservationtheta}) a linear combination of the quasi-Einstein equations. At this stage it is clear that the interaction between the two sectors is completely gravitational. It is reflected in the equations (\ref{conservde}) and (\ref{conservationtheta}), where both sectors are individually conserved.\\ 
Summarizing, we began with a complete general system of equations (\ref{effectivedensity})-(\ref{effectivetangentialpressure}). Then a linear mapping over the radial metric function is performed (\ref{deformationlambda}), which leads to two decoupled system of equations. The system corresponding to a perfect fluid sector  $\{\tilde{\rho},\tilde{p},\nu,\mu\}$ given by (\ref{ro1})-(\ref{p2}) is completely determined once we pick a well behaved isotropic solution.  To the remainnig equations (\ref{cero})-(\ref{dos}) one can imposes some constrints over the unknown functions $\{f^{*},\theta^{t}_{t},\theta^{r}_{r},\theta^{\varphi}_{\varphi}\}$ in order to generate the anisotropic solution, which it described by the thermodynamic observables (\ref{effecrho})-(\ref{effecpt}).
\section{Charged anisotropic Heintzmann solution}

The above explanation w.r.t MGD approach, is the most general case where the input corresponds to a perfect fluid solution. However, the seed can be another type of matter distribution. For example, it could be anisotropic from the beginning i.e described by 
\begin{equation}\label{anisotropicseed}
\tilde{T}_{\mu\nu}=\left(\tilde{\rho}+\tilde{p}_{r}\right)u_{\mu}u_{\nu}-\tilde{p}_{t}g_{\mu\nu}+\left(\tilde{p}_{r}-\tilde{p}_{t}\right)\eta_{\mu}\eta_{\nu},  
\end{equation}
with $u_{\mu}$ being the fluid four-velocity and $\eta_{\mu}$ a spacelike vector which is orthogonal to $u_{\mu}$. Another option is take a perfect fluid coupled to a static electric field (like in our case), where the energy-momentum tensor reads
\begin{equation}\label{perfectfluidelectric}
\begin{split}
\tilde{T}_{\mu\nu}=\left(\tilde{\rho}+\tilde{p}\right)u_{\mu}u_{\nu}-\tilde{p}g_{\mu\nu}   +\frac{1}{4\pi}\left(-F_{\mu}^{\sigma}F_{\nu\sigma}
+\frac{g_{\mu\nu}}{4}F^{\alpha\beta}F_{\alpha\beta}\right).    
\end{split}
\end{equation}
Here $F_{\mu\nu}$ is the anti-symmetric electromagnetic field tensor and can be defined as 
\begin{equation}
F_{\mu\nu}=\partial_{\mu}A_{\nu}-\partial_{\nu}A_{\mu}
\end{equation}
where $A_{\mu}=\left(\phi(r),0,0,0\right)$ is the four-potential. $F_{\mu\nu}$ satisfies the covariant Maxwell equations, given by
\begin{equation}\label{maxwell1}
\partial_{\mu}\left[\sqrt{-g}F^{\nu\mu}\right]=4\pi\sqrt{-g} J^{\nu},    
\end{equation}
\begin{equation}\label{maxwell2}
\partial_{\alpha}F_{\beta\sigma}+\partial_{\beta}F_{\sigma\alpha}+\partial_{\sigma}F_{\alpha\beta}=0,
\end{equation}
where $J^{\nu}$ is the electromagnetic four-current vector defined as
\begin{equation}\label{fourdensitycurrent}
J^{\nu}=\sigma u^{\nu},   
\end{equation}
where $\sigma=e^{\nu/2}J^{0}(r)$ represents the charge density and $g$ is the determinant of the metric (\ref{schwarzschild}), which explicitly reads
\begin{equation}
g=-e^{\nu+\lambda}r^{4}\sin^{2}\theta.    
\end{equation}
For a static spherically symmetric stellar system $J^{0}$ is the only non vanishing component of the electromagnetic
four-current $J^{\nu}$ which is a purely radial function. The only non zero
components of the electromagnetic field tensor are $F^{01}$ and $F^{10}$, which are related by $F^{01}=-F^{10}$. Being both the radial component of the electric field. From equations (\ref{maxwell1}) and (\ref{fourdensitycurrent}) the electric field $E(r)$ reads
\begin{equation}\label{electricfielddefinition}
E(r)=F^{01}(r)=\frac{4\pi}{r^{2}}e^{-(\nu+\lambda)/2}\int^{r}_{0}r^{\prime2}\sigma e^{\lambda} dr^{\prime},     
\end{equation}
and the total charge $q(r)$ is given by
\begin{equation}\label{electricchargedefinition}
q(r)={E(r)}{r^{2}}.    
\end{equation}
So, from the expressions (\ref{maxwell1}), (\ref{fourdensitycurrent}), (\ref{electricfielddefinition}) and (\ref{electricchargedefinition}) the energy-momentum tensor (\ref{perfectfluidelectric}) becomes
\begin{equation}\label{reducedperfectfluidelectric}
T_{\mu\nu}=diag\left(-\tilde{\rho}-\frac{E^{2}}{8\pi},\tilde{p}-\frac{E^{2}}{8\pi},\tilde{p}+\frac{E^{2}}{8\pi},\tilde{p}+\frac{E^{2}}{8\pi}\right).    
\end{equation}
Thus, the Einstein-Maxwell equations for (\ref{reducedperfectfluidelectric}) are
\begin{eqnarray}\label{ro11}
8\pi\tilde{\rho}+E^{2}&=&\frac{1}{r^{2}}-\frac{\mu}{r^{2}}-\frac{\mu^{\prime}}{r}\\\label{p11}
8\pi \tilde{p}-E^{2}&=&-\frac{1}{r^{2}}+\mu\left(\frac{1}{r^{2}}+\frac{\nu^{\prime}}{r}\right)\\\label{p22}
8\pi \tilde{p}+E^{2}&=&\frac{\mu}{4}\left(2\nu^{\prime\prime}+\nu^{\prime2}+2\frac{\nu^{\prime}}{r}\right)+\frac{\mu^{\prime}}{4}\left(\nu^{\prime}+\frac{2}{r}\right).
\end{eqnarray}
Now let's apply the MGD approach in order to solve the Einstein field equations for the interior of charged anisotropic compact stars. We take as a 
seed the charged Heintzmann solution  $\{\nu;\mu;\tilde{\rho};\tilde{p}\}$ modelling compact objects \cite{Thirukkanesh}. As said 
above, MGD approach decouple the system of equations (\ref{effectivedensity})
-(\ref{effectivetangentialpressure}),
one of them corresponding in this case to the  Einstein-Maxwell system (\ref{ro11})-(\ref{p22}), solved once
the seed solution is specified. In this case we have that the seed is described by
\begin{equation}\label{rhoperf}
\begin{split}
\tilde{\rho}(r)=\frac{1}{16\pi\left(1+4ar^{2}\right)^{3/2}\left(1+ar^{2}\right)^{2}}\Bigg[\left(12a^{3}r^{4}+39a^{2}r^{2}+9a\right)\left(1+4ar^{2}\right)^{1/2} &\\ +9\left(1+3ar^{2}\right)ac-2\left(32r^{4}a^{2}+46ar^{2}+11\right)\beta r^{2}\Bigg],
\end{split}
\end{equation}
\begin{equation}\label{preperf}
\begin{split}
\tilde{p}(r)=\frac{3}{16\pi\left(1+4ar^{2}\right)^{3/2}\left(1+ar^{2}\right)^{2}}\Bigg[\left(3a-3a^{2}r^{2}\right)\left(1+4ar^{2}\right)^{1/2}-\left(1+7ar^{2}\right)ca & \\ +\left(2+12r^{2}\right)\beta r^{2}\Bigg],   
\end{split}
\end{equation}
with the following metric components 
\begin{eqnarray}\label{nu}
e^{\nu(r)}&=&A^{2}\left(1+ar^{2}\right)^{3} \\ \label{mu}
\mu(r)&=&1-\frac{3ar^{2}}{2}\left[\frac{1+\left(c-\frac{4\beta r^{2}}{3a}\right)\left(1+4ar^{2}\right)^{-1/2}}{1+ar^{2}}\right],
\end{eqnarray}
which are regular everywhere inside the star even at the center $r=0$, where $e^{\lambda(r=0)}=\mu(r=0)=1$ and $e^{\nu(r=0)}> 0$. The constant parameters $A$, $a$, $c$ and $\beta$, will be determined using junction conditions at the surface $r=R$.  For this purpose the interior solution will be joined smoothly  at the surface of spheres with the exterior Reissner $-$Nordstrom solution. Here the $\beta$ parameter is related with the electric field, given by 
\begin{equation}\label{electricfield}
E^{2}(r)=\frac{\beta r^{2}\sqrt{1+4ar^{2}}}{\left(1+ar^{2}\right)^{2}}.
\end{equation}
Once the system of equations (\ref{effectivedensity})-(\ref{effectivetangentialpressure}) has been decoupled, the remaining equations (\ref{cero})-(\ref{dos}) must be solved in order to obtain an anisotropic solution. For that, it is unavoidable to choose reasonable constraints that lead to physically acceptable solutions.
The next section shows at least one restriction that leads to an admissible solution from the physical point of view.

\subsection{Mimicking the pressure for the anisotropy}

The closure of the system (\ref{effectivedensity})-(\ref{effectivetangentialpressure}) must be complemented with extra information. In principle nothing prevents us to choose some expression for $f^{*}(r)$ that results in a physically well-behaved solution, or perhaps impose some restrictions on $\theta_{\mu\nu}$ that leads to the desired result. In this opportunity we consider a restriction on $\theta^{r}_{r}$, imposing that it be equal to the pressure $\tilde{p}$ of the seed solution
\begin{equation}\label{mimicking}
\theta^{r}_{r}(r)=\tilde{p}(r).    
\end{equation}
The previous assignment establishes a direct relationship between equations (\ref{p1}) and (\ref{one}), from which the following expression is derived for $f^{*}(r)$
\begin{equation}\label{f}
f^{*}(r)=-\mu(r)+\frac{1}{1+r\nu^{\prime}(r)}.    
\end{equation}
Thus the deformed radial component (\ref{deformationlambda}) becomes to 
\begin{equation}\label{modifiedmu}
e^{-\lambda}\mapsto \left(1-\alpha\right)\mu(r)+\alpha \frac{1+ar^{2}}{1+7ar^{2}},     
\end{equation}
while the temporal component $e^{\nu}$ remains unchanged. Consequently (\ref{nu}) and (\ref{modifiedmu}) constitute the deformed solution
\begin{equation}\label{deformedsolution}
\begin{split}
ds^{2}=A^{2}\left(1+ar^{2}\right)^{3}dt^{2}-\Bigg[ \left(1-\alpha\right)\mu(r) 
+\alpha \frac{1+ar^{2}}{1+7ar^{2}}\Bigg]^{-1}dr^{2}   -r^{2}\left(d\theta^{2}+\sin^{2}\theta d\phi^{2}\right),
\end{split}
\end{equation}
where $\mu(r)$ is given by (\ref{mu}). Of course, taking $\alpha=0$ in (\ref{modifiedmu}) we recover the original solution (\ref{nu})-(\ref{mu}).

\section{Effective thermodynamic observables and mass function}
By virtue of the mimicking (\ref{mimicking}) and the expression given for $f^{*}(r)$ in (\ref{f}), and using the equations (\ref{cero})-(\ref{dos}) we obtain the following effective thermodynamic observables that characterize the fluid
\begin{equation}\label{finalpr}
{p}_{r}(\alpha;r)= \left(1-\alpha\right)\tilde{p}
\end{equation}
\begin{equation}\label{finalpt}
{p}_{t}(\alpha;r)= {p}_{r}+\frac{\alpha r^{2}}{8\pi}\Bigg[\frac{9a^{2}\left(7a^{2}r^{4}+10a r^{2}+3\right)}{\left(1+7ar^{2}\right)^{2}\left(1+ar^{2}\right)^{2}} 
-\frac{\beta\left(1+4a r^{2}\right)^{1/2}}{\left(1+ar^{2}\right)^{2}}\Bigg].
\end{equation}

From the latter equations, the anisotropy is directly computed; comparing with equation (\ref{anisotropy}) we obtain
\begin{equation}
\Pi(\alpha;r)=\frac{\alpha r^{2}}{8\pi}\Bigg[\frac{9a^{2}\left(7a^{2}r^{4}+10a r^{2}+3\right)}{\left(1+7ar^{2}\right)^{2}\left(1+ar^{2}\right)^{2}} 
-\frac{\beta\left(1+4a r^{2}\right)^{1/2}}{\left(1+ar^{2}\right)^{2}}\Bigg].
\end{equation}

One can go on computing the density
 following (\ref{effecrho}) with the temporal component of the anisotropy given by (\ref{cero})
\begin{equation}\label{finalrho}
\begin{split}
{\rho}(\alpha;r)&=\tilde{\rho}+\frac{\alpha}{16\pi}\bigg[\frac{9a\left(3ar^{2}+3-7a^{3}r^{6}-31a^{2}r^{4}\right)}{\left(1+7ar^{2}\right)^{2}\left(1+ar^{2}\right)^{2}}   +\frac{a^{2}r^{2}\left(32\beta r^{4}-27c\right)}{\left(1+4ar^{2}\right)^{3/2}\left(1+ar^{2}\right)^{2}} \\ & 
+\frac{a\left(76\beta r^{4}-9c\right)+20\beta r^{2}}{\left(1+4ar^{2}\right)^{3/2}\left(1+ar^{2}\right)^{2}}\bigg]. 
\end{split}
\end{equation} 
As we will see later, an admissible solution must satisfy some general physical requirements. However, we analyze some of them early in order to achieve the corresponding constants parameters that lead a well behaved anisotropic solution. These physical features are respect to the regularity of the effective thermodynamic observables $\tilde{\rho}$, $\tilde{p_{r}}$ and $\tilde{p_{t}}$ inside the star $(0\leq r \leq R)$. All of them must be positive and monotonically decreasing toward to the surface object. The effective central pressure and density at the interior are given by  
\begin{equation}\label{upperc}
8\pi{p}_{r}(r=0)=8\pi{p}_{t}(r=0)=\frac{3a\left(1-\alpha\right)\left(3-c\right)}{2}>0,     
\end{equation}
\begin{equation}
8\pi\rho(r=0)=\frac{9a}{2}\left(c-c\alpha+3\alpha+1\right)>0.
\end{equation}
To satisfy Zeldovich condition at the interior, ${p}_{r}/{\rho}$ at center must be $\leq1$. Therefore
\begin{equation}\label{lowerc}
\frac{\left(1-\alpha\right)\left(3-c\right)}{3\left(c-c\alpha +3\alpha +1\right)}\leq1.    
\end{equation}
On using (\ref{upperc}) and (\ref{lowerc}) we get a constraint on $c$ given as
\begin{equation}\label{constantc}
\frac{3\alpha}{\alpha-1}\leq c<3.    
\end{equation}
From (\ref{finalpr}) we obtain an upper limit $\alpha<1$. This ensures the positiveness of the effective radial pressure ${p}_{r}$ within the star. On the other hand (\ref{finalpt}) imposes a lower bound $\alpha>0$, this is so because ${p}_{t}>{p}_{r}>0$ everywhere inside the star. Moreover, we need to ensure the following statement in the surface: ${p}_{r}|_{r=R}=0$ (it determines the star size).   

\begin{figure}[H]
\centering
\includegraphics[scale=1.3]{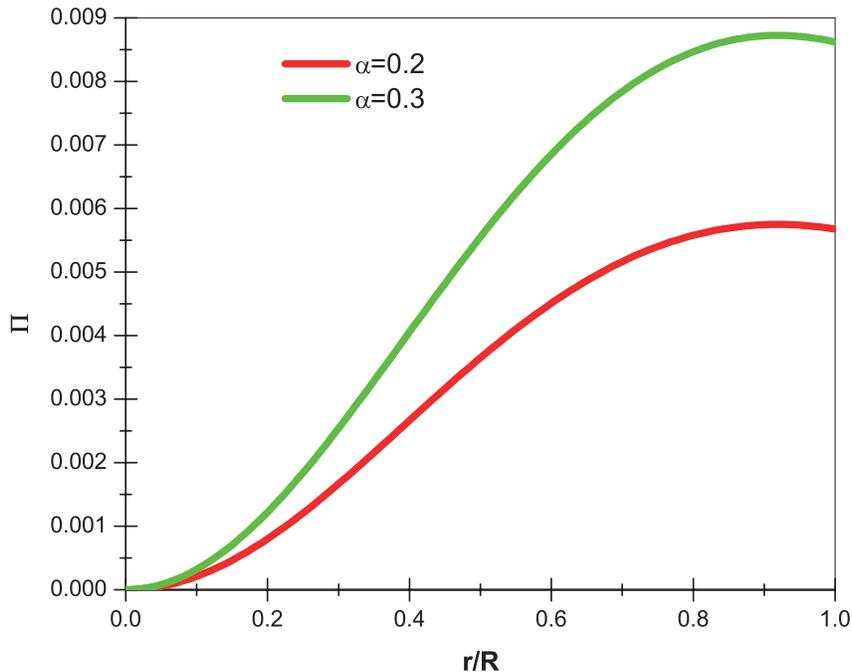}
\caption{Effective anisotropy factor $\Pi$, for the strange star candidate $RXJ 1856-37$. \label{ANI}} 
\end{figure}

It is clear from fig. (\ref{ANI}) that the  effective anisotropy $\Pi$, it vanishes at $r=0$. That is so because at the center the effective radial and transverse pressures coincide. On the other hand, as the radius increases the values of these quantities drift apart, and therefore the anisotropy increases toward the surface of the object.

\subsection{Junction conditions}
In order to generate a model of a physically realizable bounded object we need to ensure that the interior spacetime $\mathcal{M^{-}}$ must match smoothly to the exterior spacetime $\mathcal{M^{+}}$ \cite{Israel}. In our case, the interior spacetime is given by the deformed metric (\ref{deformedsolution}), and since the exterior spacetime is empty, $\mathcal{M^{+}}$ is taken to be the Reissner-Nordstrom solution
\begin{equation}
\begin{split}
ds^{2}=\left(1-\frac{2M}{r}+\frac{Q^{2}}{r^{2}}\right)dt^{2}-\left(1-\frac{2M}{r}+\frac{Q^{2}}{r^{2}}\right)^{-1}dr^{2} 
-r^{2}\left(d\theta^{2}+\sin^{2}\theta d\phi^{2}\right), \end{split}
\end{equation}

which requires the continuity of $e^{\lambda}$, $e^{\nu}$ and $q$ across the boundary $\Sigma$ (defined by $r=R$). It is known as the first fundamental form $[ds^{2}]_{\Sigma}=0$, yielding to
\begin{eqnarray} \label{contlambda}
e^{-\lambda(R)}&=&1-\frac{2\tilde{M}}{R}+\frac{Q^{2}}{R^{2}} \\ \label{contnu}
e^{\nu(R)}&=&1-\frac{2\tilde{M}}{R}+\frac{Q^{2}}{R^{2}}\\ \label{contcharge}
q(R)&=&Q.
\end{eqnarray}
On the other hand the effective radial pressure (\ref{effecpr}) vanishes at the surface star $(r=R)$, consequently 
\begin{equation}\label{second}
{p}_{r}|_{r=R^{-}}=\left(\tilde{p}-\alpha \theta^{r}_{r}\right)|_{r=R^{-}}=0.  
\end{equation}
The above expression corresponds to the second fundamental form $[G_{\mu\nu}x^{\nu}]_{\Sigma}=0$, where $x^{\nu}$ is a unit vector projected in the radial direction. Due the election (\ref{mimicking}), equation (\ref{second}) is equivalent to request $\tilde{p}(R)=0$ in (\ref{preperf}). Therefore, we obtain the following expression for the constant $\beta$
\begin{equation}
\beta = \frac{a(3\sqrt{(4R^{2}a+1)}aR^{2}+7acR^{2}-3\sqrt{(4R^{2}a+1)}+c)}{2R^{2}(6R^{2}a+1)}.    
\end{equation}
So, the remaining constants $A$ and $a$ are obtained from (\ref{contlambda}) and (\ref{contnu}), it explicitly reads
\begin{eqnarray}\label{consA}
A^{2}\left(1+aR^{2}\right)^{3}&=&1-\frac{2\tilde{M}}{R}+\frac{Q^{2}}{R^{2}}\\ \label{consa}
\left(1-\alpha\right)\mu(R)+\alpha \frac{1+aR^{2}}{1+7aR^{2}}&=&1-\frac{2\tilde{M}}{R}+\frac{Q^{2}}{R^{2}}.
\end{eqnarray}
However  in order to close the matching conditions, the parameters $\tilde{M}$ and $R$ for strange star candidates have been used \cite{thikekar}. Tables (\ref{table1}), (\ref{table3}) and (\ref{table5}) shown all the constant parameters calculated for different values of the dimensionless coupling constant $\alpha$.

\subsection{Mass function}
The mass function $m(r)$ can be calculated from
\begin{equation}
e^{-\lambda(r)}=1-\frac{2m(r)}{r}+\frac{q^{2}(r)}{r^{2}},    
\end{equation}
then using the equations (\ref{electricchargedefinition}), (\ref{mu}), (\ref{electricfield}) and (\ref{modifiedmu}) we arrive to
\begin{equation}\label{massfunc}
\begin{split}
m(r)=\frac{r}{4\sqrt{4ar^2+1}(ar^2+1)}\Bigg[\frac{3ar^2(ar^2+1)(\sqrt{4ar^2+1}+c)+2\beta r^4(2ar^2-1)}{(ar^2+1)}& \\ +
\alpha\bigg[\frac{(4+17ar^2-5a^2r^4)\sqrt{4ar^2+1}}{(7ar^2+1)}
-\frac{r^2(7ar^2+1)(3ac-4\beta r^2)}{(7ar^2+1)}\bigg]
\Bigg].   
\end{split}   
\end{equation}
We observe from (\ref{massfunc}) that $m(0)=0$. However $m^{\prime}(r)$ is positive for $r >0$. It indicates that $m(r)$ is increasing monotonically away from centre and attains regular minimum at $r=0$.

\begin{figure}[H]
\centering
\includegraphics[scale=1.3]{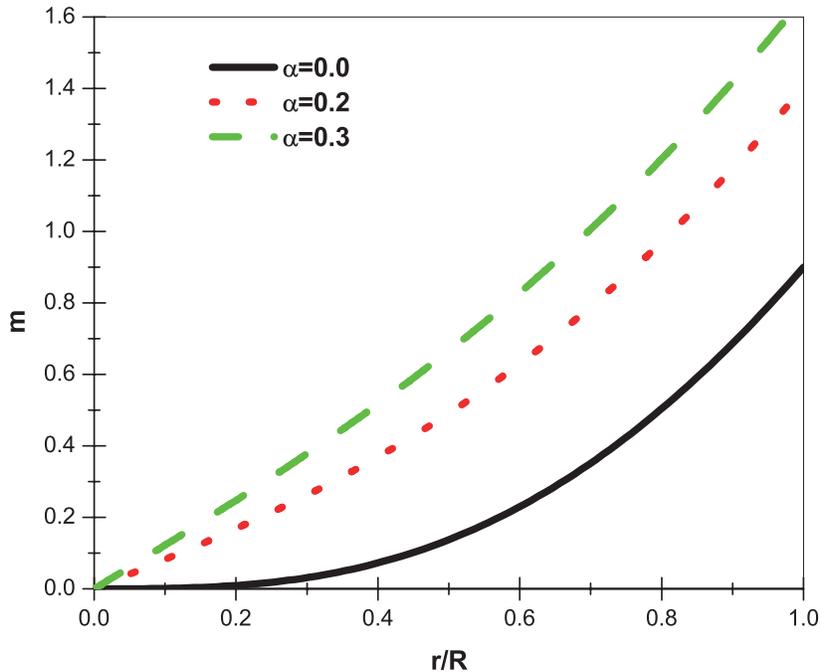}
\caption{The mass function $m(r)$ versus the fractional radius $r/R$, for the strange star candidate $RXJ 1856-37$. The solid black line corresponds to the seed solution (hereinafter), while the dotted (red line) and the dashed line (green line) are the corresponding minimal deformed metrics for $\alpha=0.2$ and $\alpha=0.3$ respectively. \label{massfunction}} 
\end{figure}

\section{Physical features}
In order to be physically meaningful, the interior solution for static fluid spheres must satisfy some more general physical requirements.
The following conditions have been generally recognized to be crucial for anisotropic fluid spheres \cite{Herrera}
\begin{enumerate}
 \item The solution should be free from physical and geometric singularities and non zero positive values of $e^{\lambda}$ and $e^{\nu}$ i.e. $(e^{\lambda})_{r=0}=1$ and $e^{\nu}>0$.
 \item The radial pressure $p_{r}$ must be vanishing but the tangential pressure $p_{t}$ may not vanish at the boundary $r=R$ of the sphere. However the radial pressure equal to the tangential pressure at the centre of the fluid sphere.
 \item The density $\rho$ and pressures $p_{r}$, $p_{t}$ should be positive inside the star.
 \item $\left(\frac{dp_{r}}{dr}\right)_{r=0}=0$ and $\left(\frac{d^{2}p_{r}}{dr^{2}}\right)_{r=0}<0$  so that
pressure gradient $\frac{dp_{r}}{dr}$ is negative for $0<r\leq R$.
 \item $\left(\frac{dp_{t}}{dr}\right)_{r=0}=0$ and $\left(\frac{d^{2}p_{t}}{dr^{2}}\right)_{r=0}< 0$ so that pressure
gradient $\frac{dp_{t}}{dr}$ is negative for $0<r\leq R$.
 \item $(\frac{d\rho}{dr})_{r=0}=0$ and $\left(\frac{d^{2}\rho}{dr^{2}}\right)_{r=0}<0$ so that density
gradient $\frac{d\rho}{dr}$ is negative for $0<r\leq R$.
The condition (4), (5) and (6) imply that pressure and density should be maximum at the centre and monotonically decreasing towards the surface.
 \item  Inside the static configuration the speed of sound should be less than the speed of light, i.e.
$0\leq\sqrt{\frac{dp_{r}}{d\rho}}<1$ and  $0\leq\sqrt{\frac{dp_{t}}{d\rho}}<1$.
In addition to the above, the velocity of sound should be decreasing towards the surface. i.e.
$\frac{d}{dr}\left(\frac{dp_{r}}{d\rho}\right)<0$
or $\left(\frac{d^{2}p_{r}}{d\rho^{2}}\right)>0$ and 
$\frac{d}{dr}\left(\frac{dp_{t}}{d\rho}\right)<0$ or $\left(\frac{d^{2}p_{t}}{d\rho^{2}}\right)>0$ for
$0\leq r \leq R$ i.e. the velocity of sound is increasing with the increase of density.
\item A physically reasonable energy-momentum tensor
has to obey the null energy condition (NEC), weak energy condition (WEC), strong energy condition (SEC) and the dominant energy condition (DEC). 
\item ) Electric intensity $E$, such that $E(0)= 0$, is taken to be monotonically increasing i.e. $(dE/dr) > 0$ for $0<r<R$.
\item The central red shift $Z_{0}$ and surface red shift $Z_{R}$ should be positive and finite i.e. $Z_{0}= \left[e^{-\nu(r)/2}-1\right]_{r=0}>0$ and $Z_{R} =\left[e^{\lambda(r)/2} - 1\right]_{r=R}>0$ and both should be bounded.
\end{enumerate}

\subsection{Regularity of the metric functions at the center}

\begin{figure}[H]
\centering
\includegraphics[width=\textwidth]{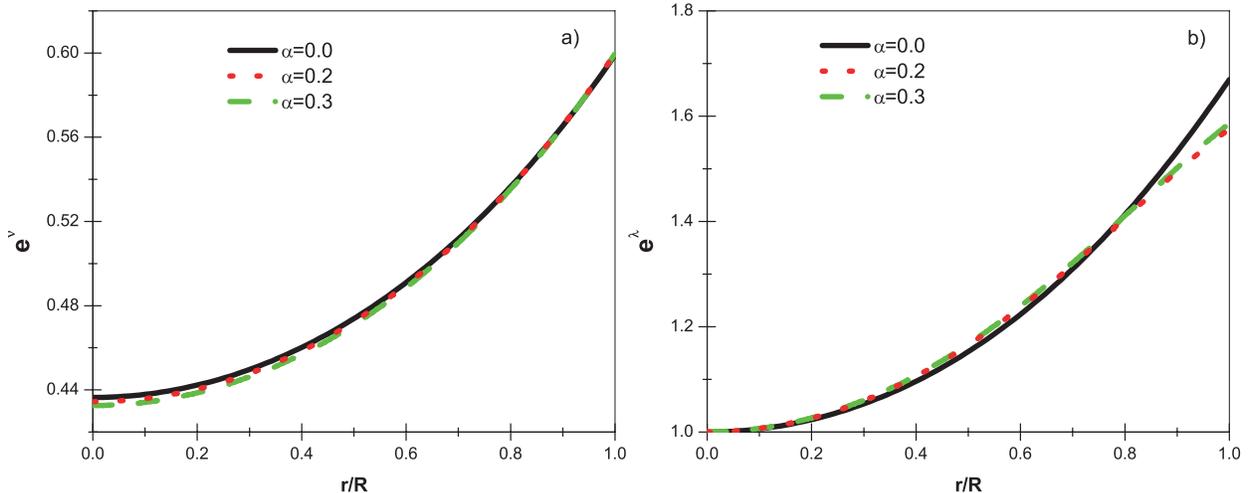}
\caption{Panel $a)$ shows the behaviour of the temporal metric function $e^{\nu(r)}$. At the center it is completely regular, finite and positive. Panel $b)$ displays the behaviour of the radial function, which is equal to $e^{\lambda(0)}=1$ at $r=0$. These plots correspond to the strange star candidate $RXJ 1856-37$. \label{metricfunctions}}
\end{figure}

A well behaved spherically symmetric and static solution of the Einstein's gravitational field equations should be free of geometric singularities. This means that the temporal $e^{\nu(r)}$ and the radial $e^{\lambda(r)}$ metric functions are continuous within the star, and completely regular at the object center $r=0$. The corresponding behaviour of the metric functions inside the compact object it shown in figure (\ref{metricfunctions}).

\subsection{Effective thermodynamic quantities}

\begin{figure}[H]
\centering
\includegraphics[width=\textwidth]{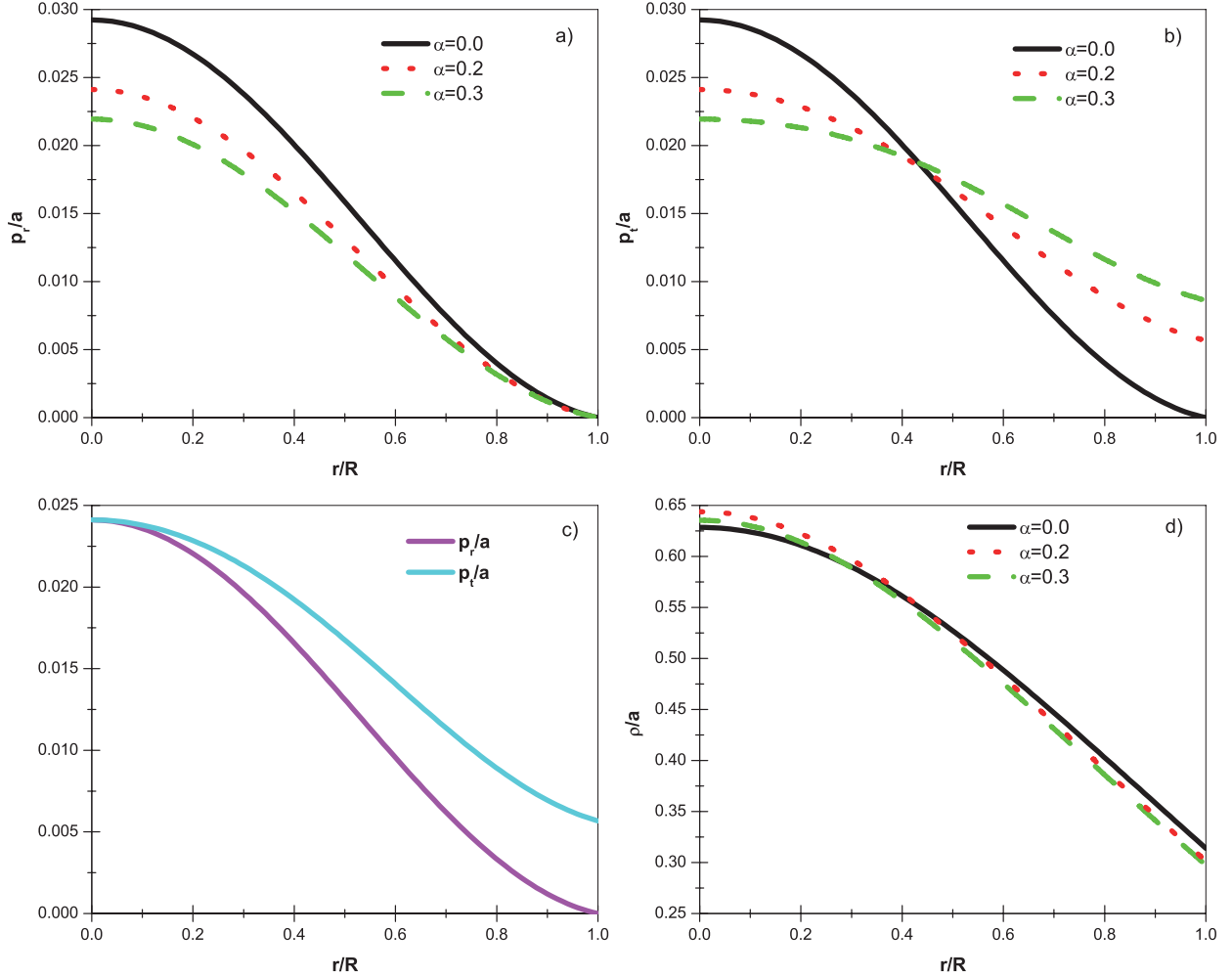}
\caption{Panels $a)$ and $b)$ show the dimensionless effective radial and tangential pressure respectively against the dimensionless radius. Panel $c)$ exhibits  a comparison between the radial and tangential pressure for $\alpha=0.2$. The anisotropy causes the pressures values to drift apart. Finally, panel $d)$ shows the dimensionless effective density energy for different values of the constant $\alpha$. All these plots correspond to the strange star candidate $RXJ1856-37$. \label{pressures}}
\end{figure}

Respect to the effective quantities, say ${p}_{r}$, ${p}_{t}$ and ${\rho}$
they must be positive, finite and monotonically decreasing towards the surface through the star. Moreover all these observables have their maximum value at the center of the object. On the other hand, the ratios $dp_{r}/d \rho$ and $dp_{t}/d \rho$ obey the Zeldovich's condition $\leq 1$. 
In the figure (\ref{pressures}) panel $c)$, is noteworthy  the presence of a force due to the anisotropic nature of the fluid. This force is directed outward when $p_{t}>p_{r}$ (inward otherwise). In this case we are in presence of a repulsive force, which allows the construction of more compact objects when using anisotropic fluid than when using isotropic fluid \cite{mehra,harko1}.

\subsection{Causality condition}
\begin{figure}[H]
\centering
\includegraphics[width=\textwidth]{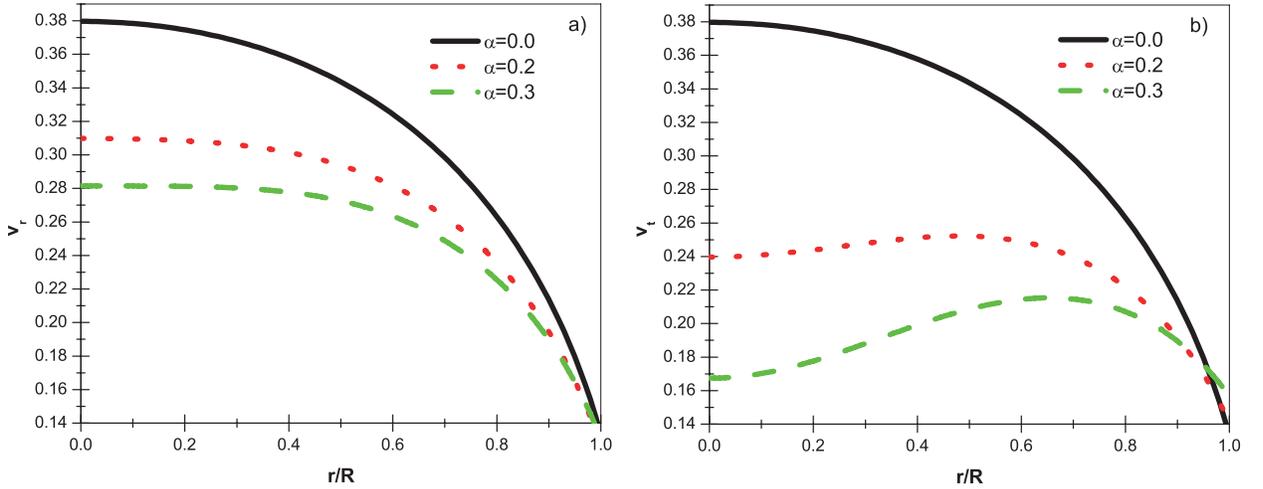}
\caption{Variation of the sound speed versus the fractional radius $r/R$ for the strange star candidate $RXJ 1856-37$. Panel $a)$ corresponds to the radial sound speed and panel $b)$ to the transverse sound speed.  \label{velocities}}
\end{figure}

The anisotropic models should satisfy the causality conditions,
i.e. $0 \leq v_{r}=\sqrt{\frac{dp_{r}}{d\rho}}<1$ and $0\leq v_{t}=\sqrt{\frac{dp_{t}}{d\rho}}<1$, at all
points inside the star. From Fig. (\ref{velocities}), we can see that our model is satisfying the above  causality conditions. Moreover, the velocities of sound $v_{r}$ and $v_{t}$ are increasing with the increase of density and it should be decreasing outwards. Therefore, we observe that the speed of sound decreases monotonically from the center of
star (high density region) towards the surface of the star (low density region). So our anisotropic solution is well behaved.

\begin{figure}[H]
\centering
\includegraphics[width=\textwidth]{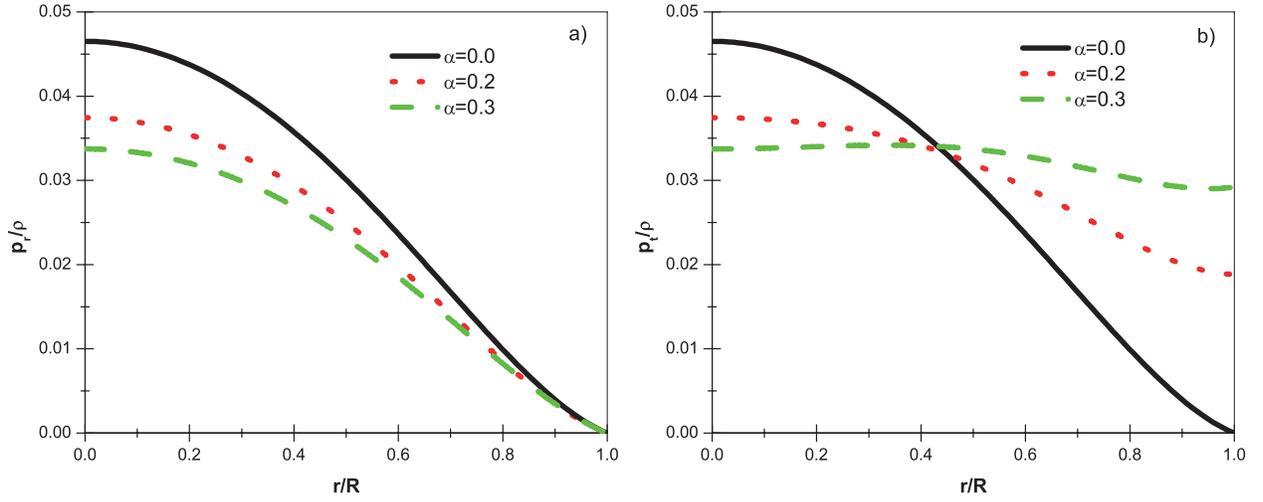}
\caption{Zeldovich's condition for the ratios $p_{r}/\rho$ (left panel) and $p_{t}/\rho$ (right panel) against the dimensionless  radius, for the strange star candidate $RXJ 1856-37$. \label{graficoheint}}
\end{figure}

\subsection{Energy conditions}

The charged anisotropic fuid sphere should satisfy the following energy conditions: (i) null energy condition (NEC), (ii) weak energy condition (WEC), (iii) strong energy condition (SEC) and (iv) dominant energy condition (DEC). For satisfying the above energy conditions, the following inequalities must be hold simultaneously inside the charged fluid sphere \citep{Leon,visserbook}
\begin{enumerate}
    \item (NEC): $\rho+p_{r} \geq 0$, $\rho+p_{t}+\frac{E^{2}}{4\pi} \geq 0$.
    \item (WEC): $\rho+\frac{E^{2}}{8\pi} \geq 0$, $\rho+p_{r} \geq 0$, $\rho+p_{t}+\frac{E^{2}}{4\pi} \geq 0$ .  
    \item (SEC): $\rho+p_{r} \geq 0$, $\rho+p_{t}+\frac{E^{2}}{4\pi} \geq 0$, \\
    $\rho+2p_{t}+p_{r}+\frac{E^{2}}{4\pi} \geq 0$.
    \item (DEC): $\rho +\frac{E^{2}}{8\pi}-|p_{r}-\frac{E^{2}}{8\pi}|\geq0$, $\rho +\frac{E^{2}}{8\pi}-|p_{t}+\frac{E^{2}}{8\pi}|\geq0$.
\end{enumerate}

\begin{figure}[H]
\centering
\includegraphics[width=\textwidth]{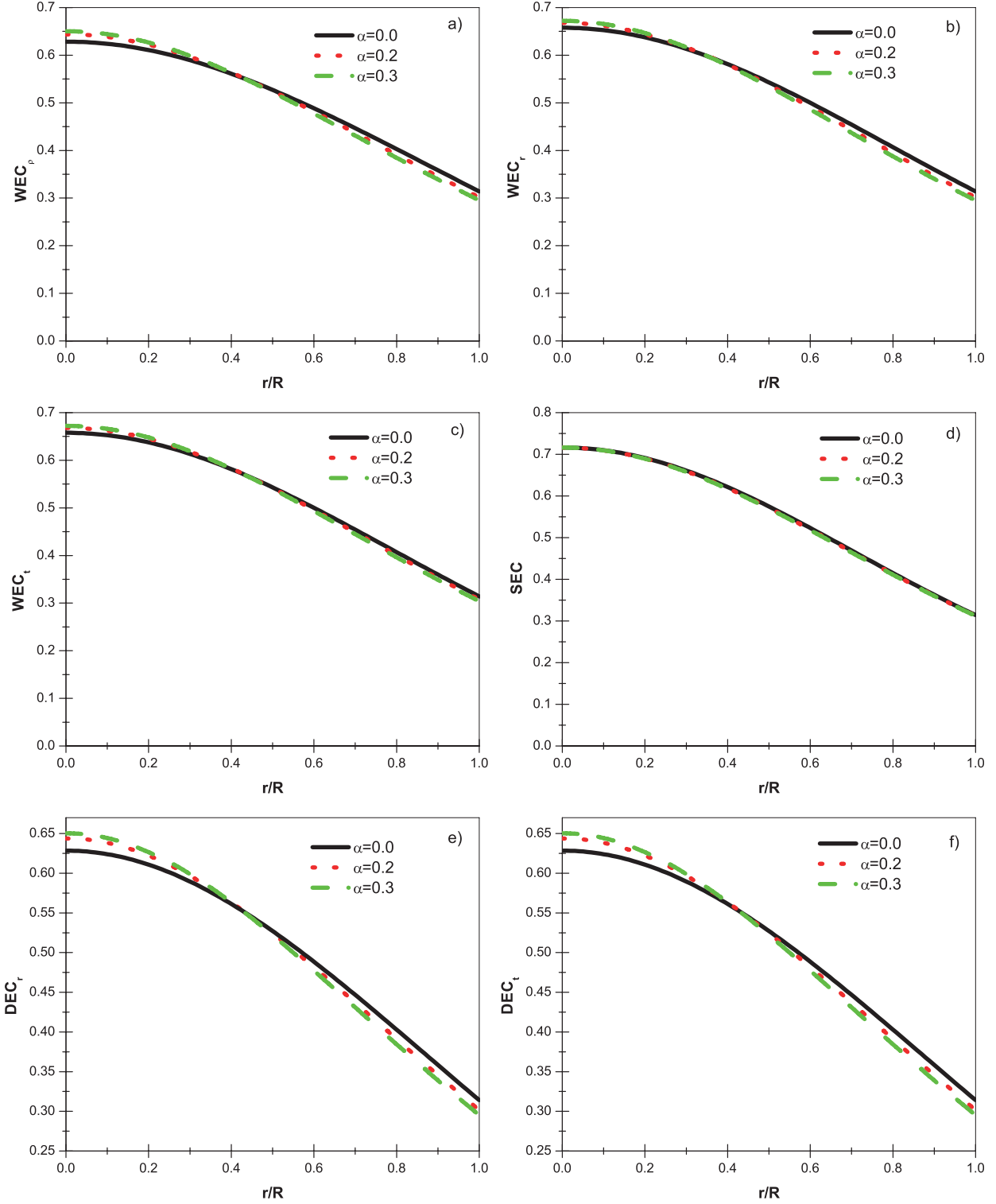}
\caption{Energy conditions for a charged anisotropic fluid sphere againts fractional radius $r/R$, corresponding to the strange star candidate $RXJ 1856-37$. \label{energyconditions}} 
\end{figure}

By continuity the (WEC) and (SEC) imply the (NEC). Figure (\ref{energyconditions}) shows that all the above inequalities are satisfied within the object. Therefore we have a well behaved energy-momentum tensor.

\subsection{Maximum allowable mass and redshift}

\begin{figure}[H]
\centering
\includegraphics[scale=1.3]{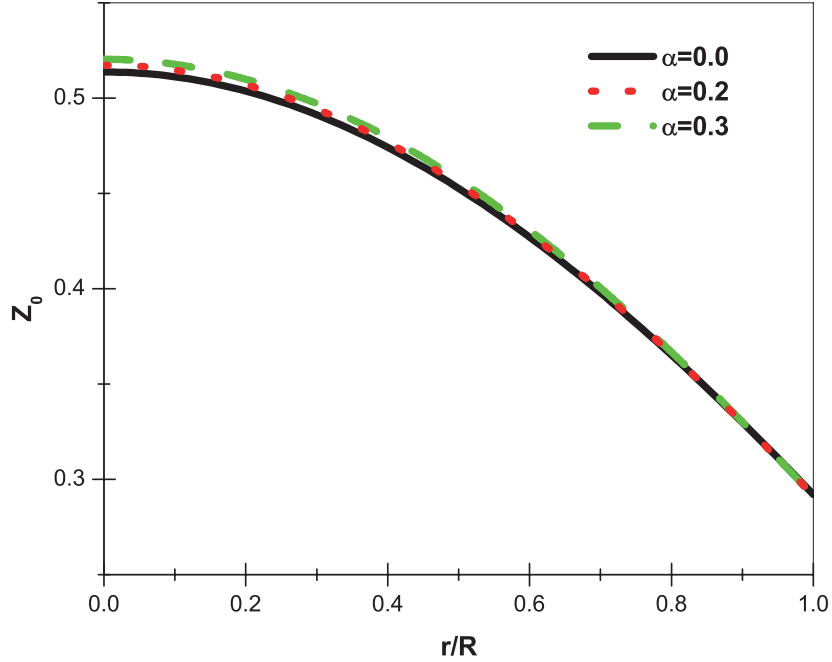}
\caption{The central redshift $Z_{0}$ against the fractional radius $r/R$, for the strange star candidate $RXJ 1856-37$. \label{central}} 
\end{figure}

A relativistic uncharged static fluid sphere has a  compactness parameter $u=M/R$ limited by $\leq 4/9$ (in the unit $c=G=1$) \cite{buchdahl}. However, the last bound has been generalized for static charged configurations. The lower limit was given by Andreasson \cite{andreason} and the upper bound was given by Boh-
mer and Harko \cite{bohmer}. This constraint on the mass-radius ratio explicitly reads
\begin{equation}
\frac{Q^{2}\left(18R^{2}+Q^{2}\right)}{2R^{2}\left(12R^{2}+Q^{2}\right)}\leq \frac{M}{R}\leq \frac{4R^{2}+3Q^{2}+2R\sqrt{R^{2}+3Q^{2}}}{9R^{2}}.
\end{equation}
So, the compactness parameter $u$, can be expresses in terms of the effective mass $M_{eff}$ which for charged matter distribution is given by \cite{maurya}
\begin{equation}
M_{eff}=4\pi\int_{0}^{R}\left({\rho}+\frac{E^{2}}{8\pi}\right)r^{2}dr=\frac{R}{2}\left[1-e^{-\lambda(R)}\right],
\end{equation}
explicitly
\begin{equation}
M_{eff}=\frac{R}{2}\Bigg[1-\alpha \frac{1+aR^{2}}{1+7aR^{2}}
-\left(1-\alpha\right)\Bigg(1-\frac{3aR^{2}}{2}\Bigg[\frac{\left(c-\frac{4\beta R^{2}}{3a}\right)\left(1+4aR^{2}\right)^{-1/2}}{1+aR^{2}}
-\frac{1}{1+aR^{2}}\Bigg]\Bigg)\Bigg].        
\end{equation}
The compactness parameter of the star is therefore
\begin{equation}
u(R)=\frac{M_{eff}}{R}=\frac{1}{2}\left[1-e^{-\lambda(R)}\right],  
\end{equation}

\begin{equation}\label{comapctness}
u(R)=\frac{1}{2}\Bigg[1-\alpha \frac{1+aR^{2}}{1+7aR^{2}}
-\left(1-\alpha\right)\Bigg(1-\frac{3aR^{2}}{2}\Bigg[\frac{\left(c-\frac{4\beta R^{2}}{3a}\right)\left(1+4aR^{2}\right)^{-1/2}}{1+aR^{2}}
-\frac{1}{1+aR^{2}}\Bigg]\Bigg)\Bigg].       
\end{equation}
The gravtitational surface redshift corresponding to above compactness $u$ (\ref{comapctness}) can be calculated as
\begin{equation}
Z_{s}=\left(1-2u\right)^{-1/2}-1.
\end{equation}
In the case of isotropic matter distribution, the maximum possible surface redshift is $Z_{s}=4.77$. On the other hand, as was pointed out by Bowers and Liang, in the presence of anisotropic matter distribution this upper bound can be exceeded \cite{bowers}. When the anisotropy parameter is positive i.e.  $(p_{t} > p_{r})$ the surface redshift is greater than its isotropic counterpart. On the other hand, the central redshift $Z_{0}$ is
\begin{equation}
Z_{0}(r)=e^{-\nu(r)/2}-1=\frac{1}{\sqrt{A^2(1+ar^2)^3}}-1.    
\end{equation}
Its monotonically decreasing behaviour inside the compact star, is shown in Fig. (\ref{central}).

\subsection{Electric properties}
\begin{figure}[H]
\centering
\includegraphics[scale=1.3]{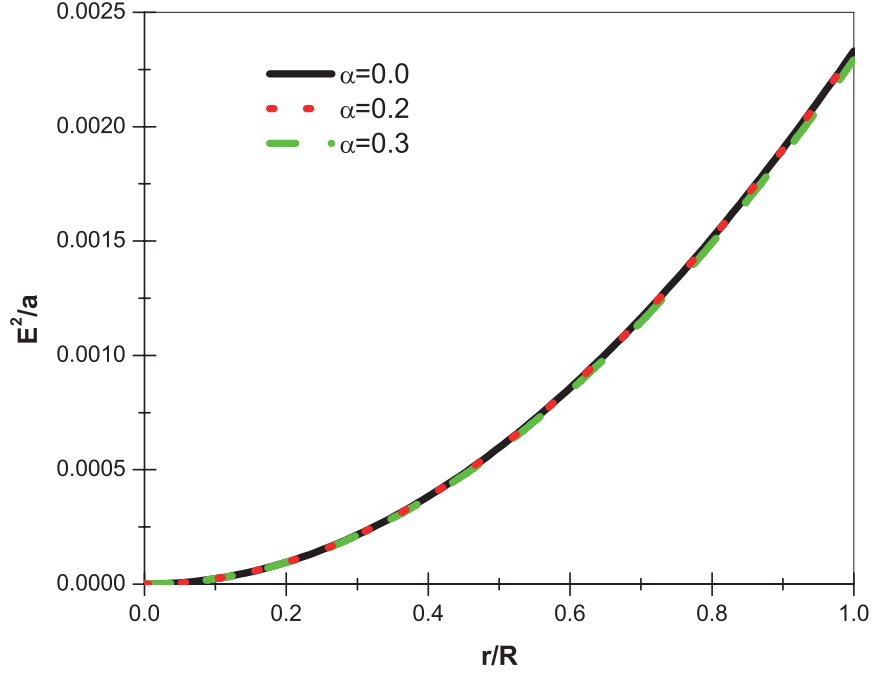}
\caption{The electric field $E$ against the fractional radius $r/R$, for the strange star candidate $RXJ 1856-37$. \label{ELEC}} 
\end{figure}

\begin{figure}[H] 
\centering
\includegraphics[width=\textwidth]{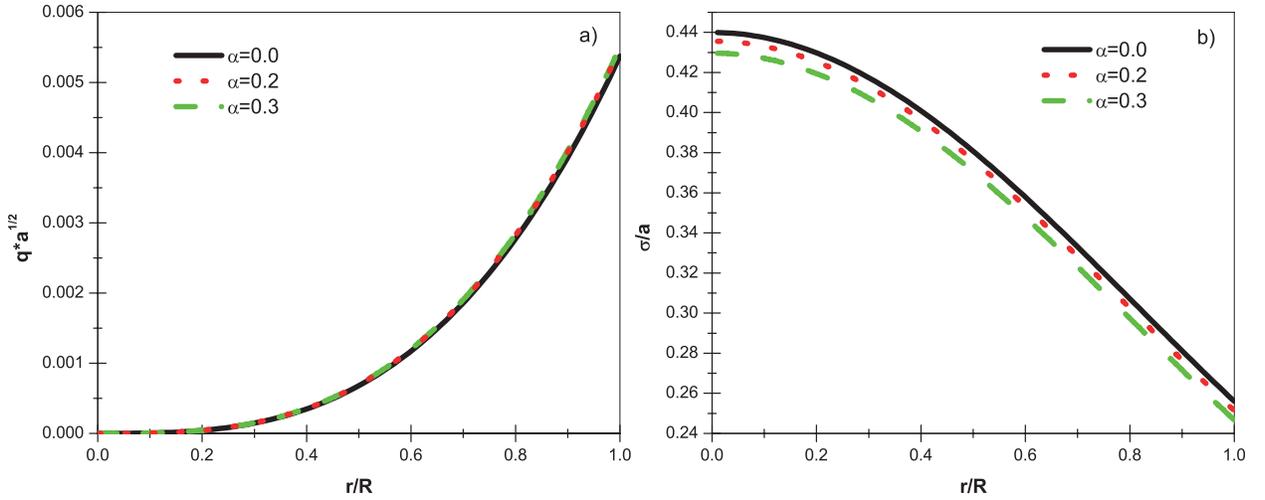}
\caption{The dimensionless electric charge (left panel) and the dimensionless charge density (right panel) versus the fractional radius $r/R$, for the strange star candidate $RXJ 1856-37$. \label{CHAR}} 
\end{figure}

We note from (\ref{electricfield}) that the electric intensity $E$ vanishes at the center of the configuration and it is monotonically increasing toward the surface of the object. The electric charged defined as  
\begin{equation}
q=Er^{2}\rightarrow q=r^{2}\sqrt{\frac{\beta r^{2}\sqrt{1+4ar^{2}}}{\left(1+ar^{2}\right)^{2}}},  
\end{equation}
has the same behaviour like the electric field $E$, i.e. null at the center and monotonically increasing with increasing radius $r$ toward the boundary of the compact star. So, the electric charge and electric field behaviour are shown in figures (\ref{CHAR}) (left panel) and (\ref{ELEC}), respectively.

On the other hand, the surface density is given by
\begin{equation}
\sigma=\frac{e^{-\lambda/2}}{4\pi r^{2}}\left(r^{2}E\right)^{\prime}.   
\end{equation}
This has its maximum in the center and decreases as it approaches to the surface of the star, as shown in Figure (\ref{CHAR}) (right panel).

\subsection{Stability conditions}

\begin{figure}[H]
\centering
\includegraphics[scale=1.3]{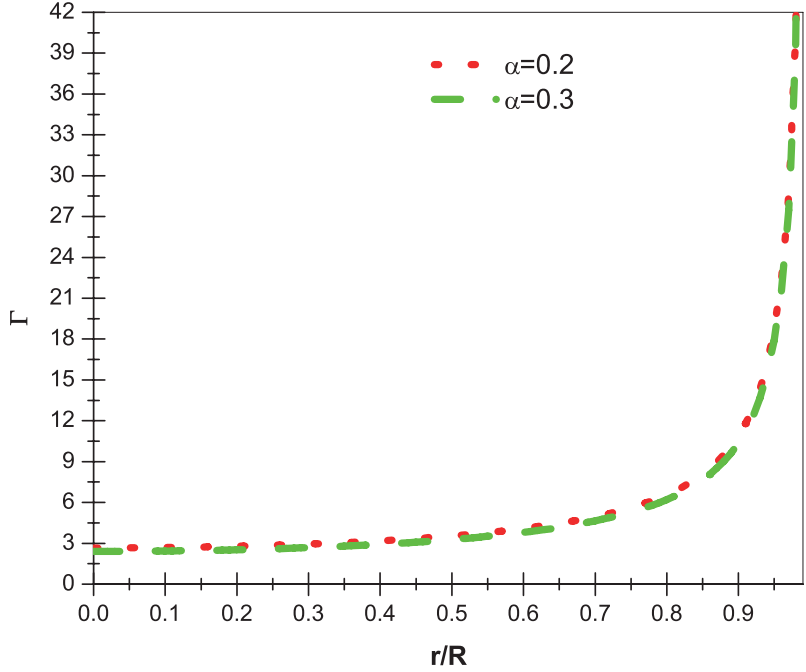}
\caption{The adiabatic index $\Gamma$ against the fractional radius $r/R$, for the strange star candidate $RXJ 1856-37$. \label{adiabaticindex}} 
\end{figure}

\begin{figure}[H]
\centering
\includegraphics[scale=1.3]{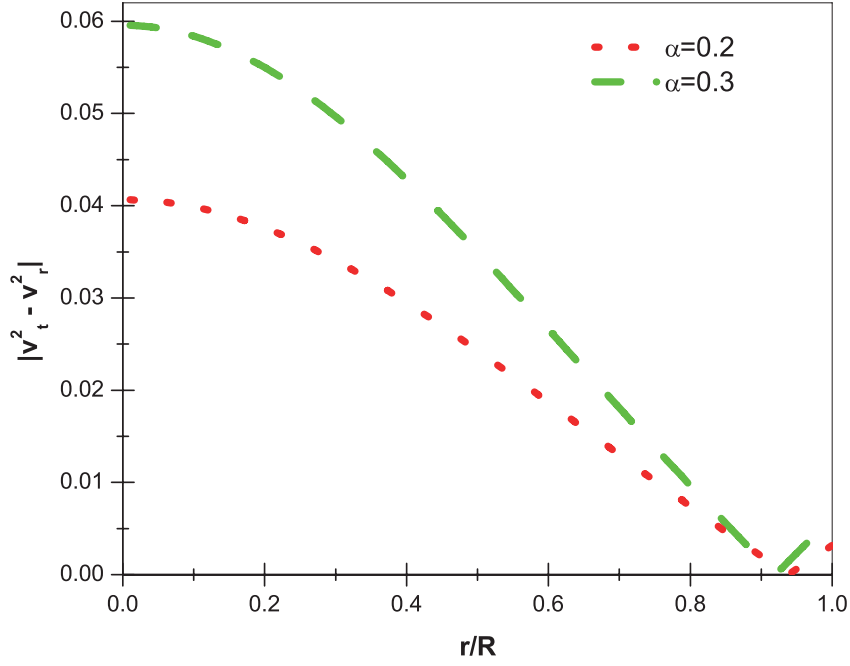}
\caption{Variation of the absolute value of square of sound velocity with respect to fractional
radius $r/R$, for the strange star candidate $RXJ 1856-37$. \label{cracking}} 
\end{figure}

\begin{figure}[H]
\centering
\includegraphics[scale=1.3]{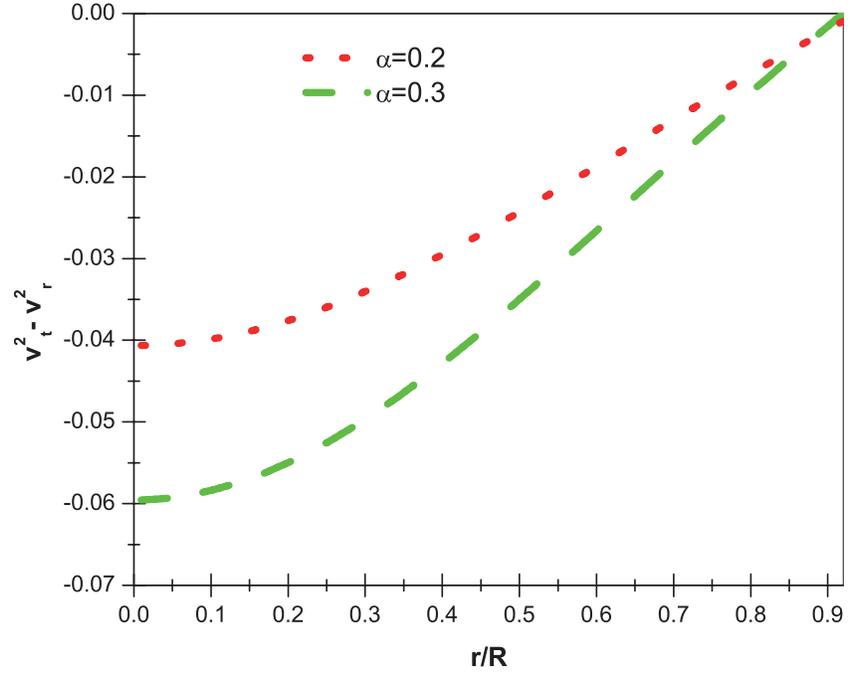}
\caption{The difference $v^{2}_{t}-v^{2}_{r}$ against the fractional radius $r/R$, for the strange star candidate $RXJ 1856-37$. \label{cracking1}} 
\end{figure}

An important analysis in the study of compact objects in GR are the stability conditions, we analyze both the adiabatic index $\Gamma$ and the radial and tangential sound velocities. In the case of a Newtonian isotropic matter distribution, it is well known that the collapsing condition corresponds to $\Gamma<4/3$ \cite{bondi,heintz}. On the other hand, with respect to relativistic anisotropic fluid spheres the above collapsing condition becomes \cite{chan1,chan2}
\begin{equation}\label{adibatic}
\Gamma<\frac{4}{3}+\left[\frac{1}{3}\kappa\frac{\rho_{0}p_{r0}}{|p^{\prime}_{r0}|}r+\frac{4}{3}\frac{\left(p_{t0}-p_{r0}\right)}{|p^{\prime}_{r0}|r}\right]_{max}    
\end{equation}
where $\rho_{0}$, $p_{r0}$ and $p_{t0}$ are the initial density, radial and tangential pressure when the fluid is in static equilibrium. The second term in the right hand side represents the relativistic corrections to the Newtonian perfect fluid and the third term is the contribution due to anisotropy. It is clear from (\ref{adibatic}) that if we have a non-relativistic perfect fluid matter distribution the bracket vanishes and we recast the collapsing Newtonian limit $\Gamma<4/3$. Heintzmann and Hillebrandt \cite{heintz} showed that in the presence of a positive an increasing anisotropy factor $\Pi=p_{t}-p_{r}>0$, the stability condition for a relativistic compact object is given by $\Gamma>4/3$, that is so because positive anisotropy factor
may slow down the growth of instability. We can explicitly obtain the adiabatic index from,  \cite{chan3}
\begin{equation}
\Gamma=\frac{\rho+p_{r}}{p_{r}}\frac{dp_{r}}{d\rho}.    
\end{equation}
Figure (\ref{adiabaticindex}) shows that $\Gamma>4/3$ everywhere within the stellar interior. Therefore our model is stable.
Another way to study stability within the framework of anisotropic compact objects in GR follows from the well known cracking concept introduced by Herrera \cite{herrera}.   Based on the Herrera's cracking concept Abreu et al. \cite{abreu} showed that potentially unstable regions within the stellar matter distribution can be identified as a function of the
difference of the radial and tangential speeds. Previously we showed that the radial and tangential speeds of our model satisfy causality condition i.e. $0\leq v_{r}<1$ and $0\leq v_{t}<1$. Moreover, one expects that the square
of the above quantities should be within the range $0\leq v^{2}_{r}\leq1$ and $0\leq v^{2}_{t}\leq1$, then we have $|v^{2}_{t}-v^{2}_{r}|\leq1$. Therefore $-1\leq v^{2}_{t}-v^{2}_{r}\leq0$ represents a potentially stable regions and $0<v^{2}_{t}-v^{2}_{r}\leq1$ a potentially unstable regions. It can be seen from figure (\ref{cracking}) that $|v^{2}_{t}-v^{2}_{r}|$ at the center lies between $0$ and $1$. On the other hand, from figure (\ref{cracking1}) it is observed that $v^{2}_{t}-v^{2}_{r}$ lies between $-1$ and $0$, thus our model is stable.

\begin{figure}[H]
\centering
\includegraphics[width=\textwidth]{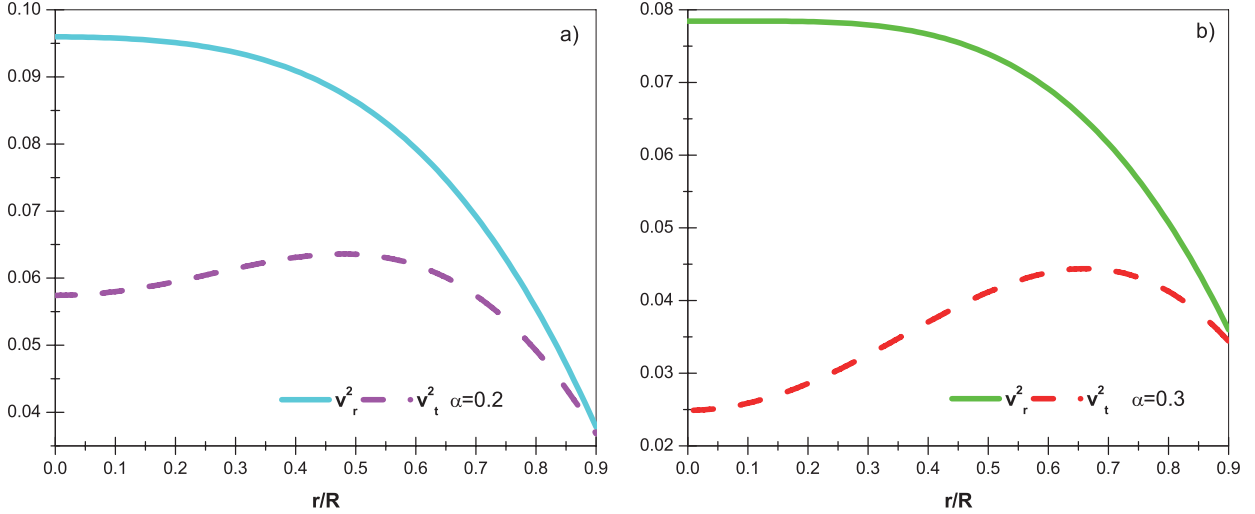}
\caption{ Variation of square of the radial and transverse velocity with respect to fractional radius $r/R$, for the strange star candidate $RXJ 1856-37$. \label{squaresvelocities}}
\end{figure}

\subsection{Equilibrium condition}
The Tolman-Oppenheimer-Volkoff
(TOV) equation for a 
char-
ged anisotropic matter fluid spheres reads \cite{maurya1}
\begin{equation}\label{TOV}
-\frac{1}{2}\nu^{\prime}\left({\rho}+{p}_{r}\right)-\frac{d{p}_{r}}{dr}+\sigma Ee^{\lambda/2}+\frac{2}{r}\left({p}_{t}-{p}_{r}\right)=0.  
\end{equation}

This equation (\ref{TOV}) describes the equilibrium condition for a charged anisotropic fluid subject to gravitational
$(F_{g})$, hydrostatic $(F_{h})$, electric $(F_{e})$ and anisotropic stress $(F_{a})$ so that
\begin{equation}
F_{g}+F_{h}+F_{e}+F_{a}=0.    
\end{equation}

\begin{figure}[H]
\centering
\includegraphics[width=\textwidth]{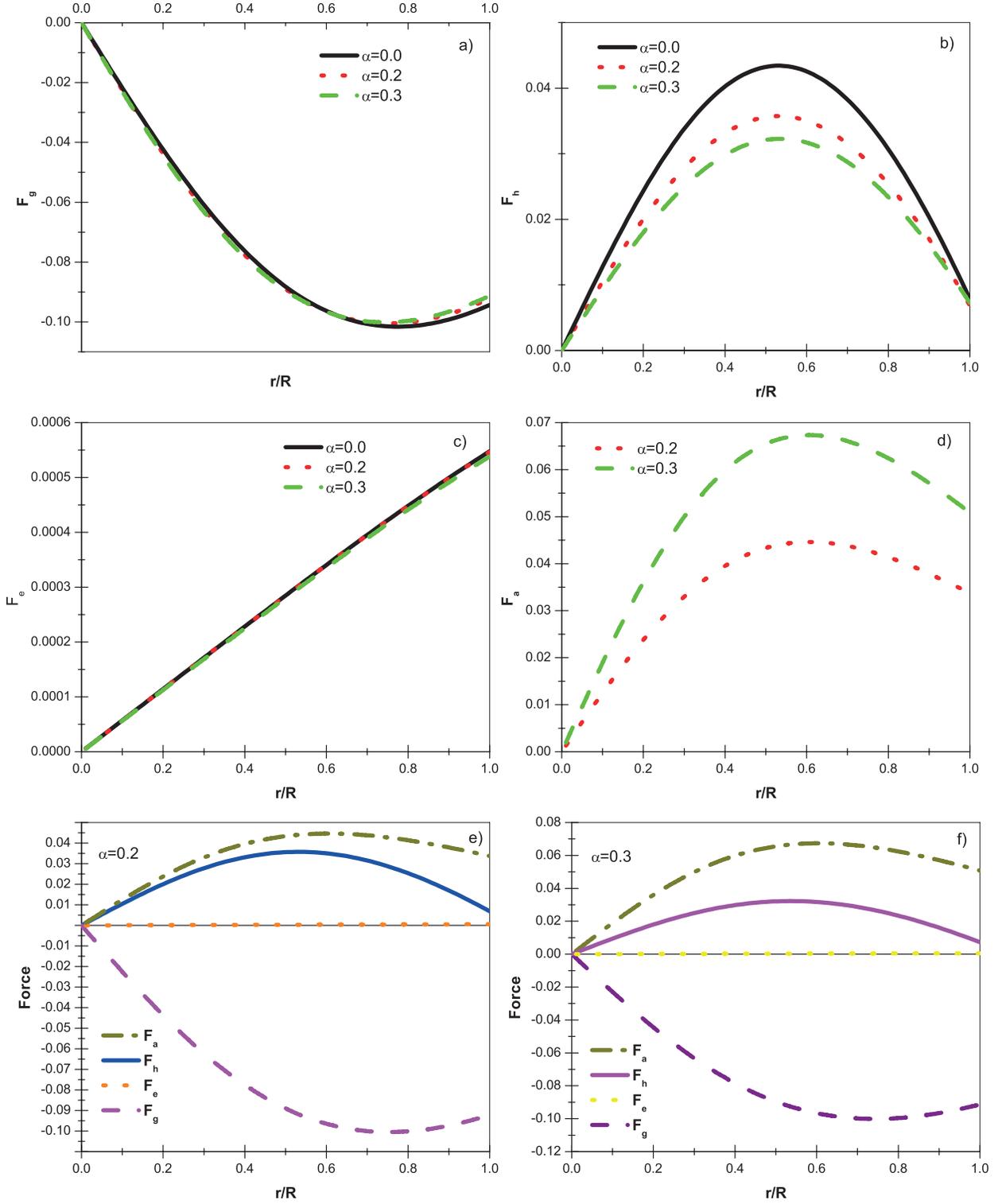}
\caption{TOV equation for static equilibrium for the strange star candidate $RXJ 1856-37$.}
\label{TOVEQ} 
\end{figure}

The figure (\ref{TOVEQ}) shows the TOV equation. It is observed that the system is in static equilibrium under four different forces, e.g. gravitational, hydrostatic, electric and anisotropic to attain overall equilibrium. However,
a strong gravitational force is counter balanced jointly by hydrostatic and anisotropic forces. Panels $e)$ and $f)$ show that the electric force, it
seems, has a negligible effect on this balancing mechanism.

To conclude the physical analysis, we summarize in tables (\ref{table2}), (\ref{table4}) and (\ref{table6}) some physical parameters, like the central and surface effective density, the central pressure, the electric field at the surface star, the surface electric charge and the ratio central pressure-central density. All these values were obtained using observational data of realistic strange star candidates e.g. $RXJ1856-37$ and $SAX J1808.4-3658$ \cite{thikekar}. 

\begin{table}[H]
\caption{Constant parameters calculated for radii and mass for some strange star candidates with $\alpha=0.0$}
\label{table1}
\begin{tabular*}{\textwidth}{@{\extracolsep{\fill}}lrrrrrl@{}}
\hline
Strange star & \multicolumn{1}{c}{radii $(R)/$} & \multicolumn{1}{c}{$M/$} & \multicolumn{1}{c}{$a/$} & \multicolumn{1}{c}{$\beta/$}&
\multicolumn{1}{c}{c (dimen-}
& A (dimen$-$ \\
candidates &$(km)$& $M_{\odot}$&$(\times10^{-3}km^{-2})$&$(\times10^{-5}km^{-4})$& sionless)& sionless) \\
\hline
RXJ 1856$-$37 & 6 & 0.9 & 3.09232 & 3.24749& 2.51019& 0.66063 \\
\hline
SAX J1808.4$-$3658 (SS2) & 6.35 & 1.323 & 4.95862 & 5.01359 & 2.08932& 0.51715 \\
\hline
\end{tabular*}
\end{table}

\begin{table}[H]
\caption{Some physical parameters calculated for radii and mass for some strange star candidates with $\alpha=0.0$}
\label{table2}
\resizebox{\textwidth}{!} {
\begin{tabular}{@{\extracolsep{\fill}}lrrrrrl@{}}
\hline
Strange star & \multicolumn{1}{c}{$\rho(0)/$} & \multicolumn{1}{c}{$\rho(R)/$} & \multicolumn{1}{c}{$p_{r}(0)/$} & \multicolumn{1}{c}{$p_{r}(0)/$}&
\multicolumn{1}{c}{$E(R)/$}
& $Q(R)/$ \\
candidates &$(\times 10^{15} gcm^{-3})$& $(\times 10^{15} gcm^{-3})$&$\rho(0)$&$(\times 10^{35} dyne/ cm^{2})$&$(\times 10^{19} Vcm^{-1})$&$(\times 10^{19} C)$\\
\hline
RXJ 1856$-$37 & 2.62243 & 1.31000 & 0.04651 & 1.09780& 3.62227& 1.44891 \\
\hline
SAX J1808.4$-$3658 (SS2) & 3.70095 & 1.36878 & 0.09826 & 3.27293 & 4.66014& 2.08787 \\
\hline
\end{tabular}
}
\end{table}

\begin{table}[H]
\caption{Constant parameters calculated for radii and mass for some strange star candidates with $\alpha=0.2$}
\label{table3}
\begin{tabular*}{\textwidth}{@{\extracolsep{\fill}}lrrrrrl@{}}
\hline
Strange star & \multicolumn{1}{c}{radii $(R)/$} & \multicolumn{1}{c}{$M/$} & \multicolumn{1}{c}{$a/$} & \multicolumn{1}{c}{$\beta/$}&
\multicolumn{1}{c}{c (dimen-}
& A (dimen$-$ \\
candidates &$(km)$& $M_{\odot}$&$(\times10^{-3}km^{-2})$&$(\times10^{-5}km^{-4})$& sionless)& sionless) \\
\hline
RXJ 1856$-$37 & 6 & 0.9 & 3.15327 & 3.31036& 2.49500& 0.65908 \\
\hline
SAX J1808.4$-$3658 (SS2) & 6.35 & 1.323 & 5.48604 & 5.49469 & 1.86200& 0.49895 \\
\hline
\end{tabular*}
\end{table}

\begin{table}[H]
\caption{Some physical parameters calculated for radii and mass for some strange star candidates with $\alpha=0.2$}
\label{table4}
\resizebox{\textwidth}{!} {
\begin{tabular}{@{\extracolsep{\fill}}lrrrrrl@{}}
\hline
Strange star & \multicolumn{1}{c}{$\rho(0)/$} & \multicolumn{1}{c}{$\rho(R)/$} & \multicolumn{1}{c}{$p_{r}(0)/$} & \multicolumn{1}{c}{$p_{r}(0)/$}&
\multicolumn{1}{c}{$E(R)/$}
& $Q(R)/$ \\
candidates &$(\times 10^{15} gcm^{-3})$& $(\times 10^{15} gcm^{-3})$&$\rho(0)$&$(\times 10^{35} dyne/ cm^{2})$&$(\times 10^{19} Vcm^{-1})$&$(\times 10^{19} C)$\\
\hline
RXJ 1856$-$37 & 2.73949 & 1.28302 & 0.03745 & 0.92332& 3.65549& 1.46219 \\
\hline
SAX J1808.4$-$3658 (SS2) & 4.09497 & 1.38541 & 0.09822 & 3.61994 & 4.38590& 1.96500 \\
\hline
\end{tabular}
}
\end{table}

\begin{table}[H]
\caption{Constant parameters calculated for radii and mass for some strange star candidates with $\alpha=0.3$}
\label{table5}
\begin{tabular*}{\textwidth}{@{\extracolsep{\fill}}lrrrrrl@{}}
\hline
Strange star & \multicolumn{1}{c}{radii $(R)/$} & \multicolumn{1}{c}{$M/$} & \multicolumn{1}{c}{$a/$} & \multicolumn{1}{c}{$\beta/$}&
\multicolumn{1}{c}{c (dimen-}
& A (dimen$-$ \\
candidates &$(km)$& $M_{\odot}$&$(\times10^{-3}km^{-2})$&$(\times10^{-5}km^{-4})$& sionless)& sionless) \\
\hline
RXJ 1856$-$37 & 6 & 0.9 & 3.19500 & 3.30625& 2.47478& 0.65771 \\
\hline
SAX J1808.4$-$3658 (SS2) & 6.35 & 1.323 & 5.10875 & 5.30252 & 2.08381& 0.51554 \\
\hline
\end{tabular*}
\end{table}

\begin{table}[H]
\caption{Some physical parameters calculated for radii and mass for some strange star candidates with $\alpha=0.3$}
\label{table6}
\resizebox{\textwidth}{!} {
\begin{tabular}{@{\extracolsep{\fill}}lrrrrrl@{}}
\hline
Strange star & \multicolumn{1}{c}{$\rho(0)/$} & \multicolumn{1}{c}{$\rho(R)/$} & \multicolumn{1}{c}{$p_{r}(0)/$} & \multicolumn{1}{c}{$p_{r}(0)/$}&
\multicolumn{1}{c}{$E(R)/$}
& $Q(R)/$ \\
candidates &$(\times 10^{15} gcm^{-3})$& $(\times 10^{15} gcm^{-3})$&$\rho(0)$&$(\times 10^{35} dyne/ cm^{2})$&$(\times 10^{19} Vcm^{-1})$&$(\times 10^{19} C)$\\
\hline
RXJ 1856$-$37 & 2.80380 & 1.27232 & 0.03374 & 0.85137& 3.65206& 1.46082 \\
\hline
SAX J1808.4$-$3658 (SS2) & 4.14544 & 1.29396 & 0.06365 & 2.37470 & 4.78444& 2.14356 \\
\hline
\end{tabular}
}
\end{table}

\section{Concluding remarks}

Gravitational decoupling through MGD is a novel approach which provides us a new branch to study self-gravitating systems with anisotropic matter distribution. In this opportunity, we have reported radial deformations only, but deformations on the temporal component of the seed metric may bring interesting results. Once the system of equations (\ref{effectivedensity})-(\ref{effectivetangentialpressure}) is decoupled, the gravitational interaction between both, the Einstein and the quasi-Einstein sectors is purely gravitational, i.e. there is no exchange of energy-momentum between them. Among all the possibilities that the method presents to solve the system of quasi-Einstein equations (\ref{cero})-(\ref{dos}), for the sake of simplicity we have chosen a simple relation between $\tilde{p}$ and $\theta^{r}_{r}$. Obtaining a well-behaved compact object model from the physical point of view. 

Particularly, we have extended the charged isotropic He-
intzmann solution to an anisotropic scenario. The resulting model fulfill all the basic criterion demanded for a well behaved solution in this context, such as: regularity of the gravitational potentials at the object center, positive definiteness and monotonic decrease behaviour  of the energy density, radial and tangential pressures with increasing radius, vanishing radial pressure at the surface star, the continuity of electric field across the boundary, the speed of sound being less than the speed of light, stability and equilibrium conditions, etc. On the other hand as was pointed out early, the presence of the electric field and the effective anisotropy counterbalance the gravitational force. In the first case due to electric repulsive force and in the second case due to repulsive gravitational force. This fact avoid the collapse of a spherically symmetric matter distribution to a point singularity during the gravitational collapse or during an
accretion process onto compact object. Moreover, in view of comparing our model with observational data of realistic stars, several physical parameters were calculated by fixing the radii and mass corresponding to the strange star candidates RXJ 1856-37 and SAX J1808.4-3658 (SS2). 

\begin{acknowledgements}
We thanks to Claudia Alvarez for useful comments and discussion. F. Tello-Ortiz thanks the financial support of the project ANT-1755 at the Universidad de Antofagasta-Chile.
\end{acknowledgements}



\begin{thebibliography}{1}

\bibitem {schwar} K. Schwarzschild, Sitz. Deut. Akad. Wiss. Berlin, Kl. Math. Phys. \textbf{24}, 424 (1916).

\bibitem{tolman} R. C. Tolman,
Phys. Rev. \textbf{55}, 364 (1939).

\bibitem{lamei} G. Lemaitre, Ann. Soc. Sci. Bruxelles A \textbf{53}, 51 (1933).

\bibitem{bowers} R. L. Bowers, E. P. T. Liang, Astrophys. J. \textbf{188}, 657 (1974).

\bibitem{ruderman} R. Ruderman, Ann. Rev. Astron. Astrophys. \textbf{10}, 427 (1972).

\bibitem{kileba} Kileba Matondo, S. D. Maharaj, S. Ray, Eur. Phys. J. C \textbf{78}, 437 (2018).

\bibitem{maurya2} S. K. Maurya, S. D. Maharaj, Eur. Phys. J. C \textbf{77}, 328 (2017).

\bibitem{hassan} Mohammad Hassan Murad, Astrophys. Space Sci. \textbf{20}, 361 (2016).

\bibitem{bhar} Pilayi Bhar, S. K. Maurya, Y. K. Gupta, Tuhina Manna, Eur. Phys. J. A \textbf{52}, 312 (2016).

\bibitem{bhar1} Pilayi Bhar, Mohammad Hassan Murad, Neeraj Pant, Astrophys. Space Sci. \textbf{13}, 359 (2015).

\bibitem{maurya3} S. K. Maurya, Y. K. Gupta, Saibal Ray, Baiju Dayanandan, Eur. Phys. J. C \textbf{75}, 225 (2015).

\bibitem{harko1} M. K. Mak, T. Harko,  Proc.Roy.Soc.Lond. A \textbf{459}, 393 (2003).

\bibitem{reissner} H. Reissner, Annalen der Physik, \textbf{50}, 106 (1916).

\bibitem{nordstrom} G. Nordstrom, Verhandl. Koninkl. Ned. Akad. Wetenschap., Afdel. Natuurk., Amsterdam. \textbf{26}, 1201 (1918).

\bibitem{Thirukkanesh} S. Thirukkanesh, F.C. Ragel, Chin. Phys. C \textbf{40}, 045101 (2016).

\bibitem{krasinski}  A, Krasinski, \emph{Inhomogeneous Cosmological Models}, (Cambridge University
Press, Cambridge, 1997).

\bibitem{bekenstein} J. D. Bekenstein, Phys. Rev. D \textbf{4}, 2185 (1971).

\bibitem{deb} Debabrata Deb, Maxim Khlopov, Farook Rahaman, Saibal Ray, B. K. Guha, Eur. Phys. J. C \textbf{78}, 465 (2018).

\bibitem{monadi} H. Panahi, R. Monadi, I. Eghdami, Chin. Phys. Lett. \textbf{33}, 072601 (2016).

\bibitem{rahaman} F. Rahaman, R. Maulick , A.K. Yadav, S. Ray, R. Sharma, Gen. Relativ. Gravit. \textbf{44}, 107 (2012).

\bibitem{varela} V. Varela, F. Rahaman, S. Ray, K. Chakraborty, M. Kalam, Phys. Rev. D \textbf{82}, 044052 (2010).

\bibitem{negreiros} R.P. Negreiros, F. Weber, M. Malheiro, V. Usov, Phys. Rev. D \textbf{80}, 083006 (2009).

\bibitem{mello} B. B. Siffert, J. R. de Mello, M. O. Calvao, Braz. J. Phys. \textbf{37}, 2B (2007).

\bibitem{ray} S. Ray, A. L. Esp\'indola, M. Malheiro, J.P.S. Lemos, V.T. Zanchin, Phys. Rev. D \textbf{68}, 084004 (2003).

\bibitem{bonnor} B. W. Bonnor, Z. Phys. \textbf{160}, 59 (1960).

\bibitem{kileba1} S. D. Maharaj, D. Kileba Matondo, P. Mafa Takisa, Int. J. Mod. Phys. D \textbf{26}, 1750014 (2017).

\bibitem{hassan1} Mohammad Hassan Murad, Saba Fatema, Int. J. Theor. Phys. \textbf{52}, 4342 (2013).

\bibitem{maurya4} S. K. Maurya, Y. K. Gupta, Astrophys. Space Sci. \textbf{344}, 243 (2013).

\bibitem{Ovalle} J. Ovalle, Phys. Rev. D \textbf{95}, 104019 (2017).

\bibitem{Ovalle9} J. Ovalle, R. Casadio, R. da Rocha, A. Sotomayor, Eur. Phys. J. C \textbf{78}, 122 (2018).  

\bibitem{randall} L. Randall, R. Sundrum,  Phys. Rev.
Lett. \textbf{83}, 3370 (1999).

\bibitem{randall1} L. Randall, R. Sundrum,  Phys. Rev. Lett. \textbf{83}, 4690 (1999). 

\bibitem{Ovalle4} R. Casadio, J Ovalle, R. da Rocha, Class. Quantum Grav. \textbf{32}, 215020 (2015). 

\bibitem{Ovalle5} J. Ovalle,  Int. J. Mod. Phys.
Conf. Ser. \textbf{41}, 1660132 (2016). 

\bibitem{Visser} P. Boonserm, T. Ngampitipan, M. Visser, Int. J. Mod. Phys. D, 1650019 (2015).

\bibitem{Ovalle1} J. Ovalle, 
Mod.Phys.Lett. A \textbf{23}, 3247 (2008).

\bibitem{Ovalle2} J. Ovalle, (ICGA9), Ed. J. Luo, World Scientific, Singapore, 173 (2010).

\bibitem{Ovalle3} J. Ovalle, F. Linares,  Phys.Rev. D \textbf{88}, no.10, 104026
(2013). 

\bibitem{Ovalle6} J. Ovalle, Laszl\'o A. Gergely, R. Casadio, Class. Quantum Grav. \textbf{32}, 045015 (2015).

 \bibitem{Ovalle7} R. Casadio, J Ovalle, R. da Rocha, Europhys. Lett. \textbf{110}, 40003 (2015). 

\bibitem{Ovalle8} J. Ovalle, R. Casadio, A. Sotomayor, Adv. High Energy Phys. \textbf{2017}, (2017).

\bibitem{Gabbanelli}  L. Gabbanelli, A. Rinc\'on, C. Rubio, Eur. Phys. J. C \textbf{78}, 370 (2018).   

\bibitem{Tello} M. Estrada, F. Tello-ortiz, (2018) arXiv:1803.02344v2 [gr-qc].

\bibitem{sharif} M. Sharif, Sobia Sadiq, Eur. Phys. J. C \textbf{78}, 122 (2018).

\bibitem{Ovalle:2018umz}
  J. Ovalle, R. Casadio, R. da Rocha, A. Sotomayor, Z. Stuchlik, (2018) arXiv:1804.03468v2 [gr-qc].

\bibitem{Israel} W. Israel, Nuovo Cim. B \textbf{44}, 1 (1966).

\bibitem{thikekar} R. Tikekar, K. Jotania,   Pramana J. Phys. \textbf{68}, 397 (2007).

\bibitem{Herrera} L. Herrera, N. O. Santos,  
Phys. Rep. \textbf{286}, 53 (1997).

\bibitem{mehra} M. K. Gokhroo, A. L. Mehra,  Gen. Rel. Grav. \textbf{26}, 75 (1994).

\bibitem{Leon} J. Ponce de Leon,  Gen. Relat. Gravit. \textbf{25}, 1123 (1993).

\bibitem{visserbook} M. Visser, \emph{Lorentzian Wormholes}, (Springer, Berlin, 1996).

\bibitem{buchdahl} H. A. Buchdahl, Phys. Rev. D \textbf{116}, 1027 (1959).

\bibitem{andreason} H. Andreasson, Commun. Math. Phys. \textbf{288}, 715 (2009).

\bibitem{bohmer} C.G. Bohmer, T. Harko, Class. Quantum Gravit. \textbf{23}, 6479 (2006).

\bibitem{maurya} S. K. Maurya, M. Govender,  Eur. Phys. J. C  \textbf{77}, 347 (2017).

\bibitem{bondi} H. Bondi, Mon. Not. R. Astron. Soc. \textbf{281}, 39 (1964).

\bibitem{heintz} H. Heintzmann, W. Hillebrandt, Astron. Astrophys. \textbf{38}, 51 (1975).

\bibitem{chan1} R. Chan, L. Herrera, N.O. Santos, Class. Quantum Grav. \textbf{9}, 133 (1992).

\bibitem{chan2} R. Chan, L. Herrera, N.O. Santos, Mon. Not. R. Astron. Soc. \textbf{265}, 533 (1993).

\bibitem{chan3} R. Chan, S. Kichenassamy, G. Le Denmat, N.O. Santos, Mon. Not. R. Astron. Soc. \textbf{239}, 91 (1989).

\bibitem{herrera} L. Herrera, Phys. Lett. A \textbf{165}, 206 (1992).

\bibitem{abreu} H. Abreu, H. Hern\'andez, L. A. N\'u\~nez, Calss. Quantum. Grav.\textbf{24}, 4631 (2007).

\bibitem{maurya1} S. K. Maurya, M. Govender, Eur.Phys.J. C \textbf{77}, 420 (2017).

\end{thebibliography}
\end{document}